\documentclass[iop,apj,twocolappendix]{emulateapj}
\usepackage{amsmath}
\usepackage{apjfonts}
\usepackage{color}
\usepackage[nolist]{acronym}
\usepackage[colorlinks,citecolor=blue,linkcolor=blue,urlcolor=blue]{hyperref}
\newif\ifinthesis
\inthesisfalse

\newcommand{\noprint}[1]{}
\newcommand{\figsetstart}{{\bf Fig. Set} }
\newcommand{\figsetend}{}
\newcommand{\figsetgrpstart}{}
\newcommand{\figsetgrpend}{}
\newcommand{\figsetnum}[1]{{\bf #1.}}
\newcommand{\figsettitle}[1]{ {\bf #1} }
\newcommand{\figsetgrpnum}[1]{\noprint{#1}}
\newcommand{\figsetgrptitle}[1]{\noprint{#1}}
\newcommand{\figsetplot}[1]{\noprint{#1}}
\newcommand{\figsetgrpnote}[1]{\noprint{#1}}

\begin{document}

\title{The Needle in the 100\,\lowercase{deg}$^2$ Haystack: Uncovering Afterglows of \emph{Fermi} GRB\lowercase{s} With the Palomar Transient Factory}

\slugcomment{The Astrophysical Journal, 806:52}
\received{2015 January 2}
\accepted{2015 February 28}
\published{2015 June 8}

\author{Leo P. Singer\altaffilmark{1,32,2}}
\email{leo.p.singer@nasa.gov}
\author{Mansi M. Kasliwal\altaffilmark{3}}
\author{S. Bradley Cenko\altaffilmark{2,4}}
\author{Daniel A. Perley\altaffilmark{33,5}}
\author{Gemma E. Anderson\altaffilmark{6,7}}
\author{G. C. Anupama\altaffilmark{8}}
\author{Iair Arcavi\altaffilmark{9,10}}
\author{Varun Bhalerao\altaffilmark{11}}
\author{Brian D. Bue\altaffilmark{12}}
\author{Yi Cao\altaffilmark{5}}
\author{Valerie Connaughton\altaffilmark{13}}
\author{Alessandra Corsi\altaffilmark{14}}
\author{Antonino Cucchiara\altaffilmark{32,2}}
\author{Rob P. Fender\altaffilmark{6,7}}
\author{Derek B. Fox\altaffilmark{15}}
\author{Neil Gehrels\altaffilmark{2}}
\author{Adam Goldstein\altaffilmark{32,16}}
\author{J. Gorosabel\altaffilmark{17,18,19}}
\author{Assaf Horesh\altaffilmark{20}}
\author{Kevin Hurley\altaffilmark{21}}
\author{Joel Johansson\altaffilmark{22}}
\author{D. A. Kann\altaffilmark{23,24}}
\author{Chryssa Kouveliotou\altaffilmark{16}}
\author{Kuiyun Huang\altaffilmark{25}}
\author{S. R. Kulkarni\altaffilmark{5}}
\author{Frank Masci\altaffilmark{26}}
\author{Peter Nugent\altaffilmark{27,28}}
\author{Arne Rau\altaffilmark{24}}
\author{Umaa D. Rebbapragada\altaffilmark{12}}
\author{Tim D. Staley\altaffilmark{6,7}}
\author{Dmitry Svinkin\altaffilmark{29}}
\author{C. C. Th\"one\altaffilmark{17}}
\author{A. de Ugarte Postigo\altaffilmark{17,30}}
\author{Yuji Urata\altaffilmark{31}}
\author{Alan Weinstein\altaffilmark{1}}

\altaffiltext{1}{LIGO Laboratory, California Institute of Technology, Pasadena, CA 91125, USA}
\altaffiltext{2}{Astrophysics Science Division, NASA Goddard Space Flight Center, Code 661, Greenbelt, MD 20771, USA}
\altaffiltext{3}{Observatories of the Carnegie Institution for Science, 813 Santa Barbara Street, Pasadena CA 91101, USA}
\altaffiltext{4}{Joint Space-Science Institute, University of Maryland, College Park, MD 20742, USA}
\altaffiltext{5}{Cahill Center for Astrophysics, California Institute of Technology, Pasadena, CA 91125, USA}
\altaffiltext{6}{Astrophysics, Department of Physics, University of Oxford, Keble Road, Oxford OX1 3RH, UK}
\altaffiltext{7}{Physics \& Astronomy, University of Southampton, Southampton SO17 1BJ, UK}
\altaffiltext{8}{Indian Institute of Astrophysics, Koramangala, Bangalore 560 034, India}
\altaffiltext{9}{Las Cumbres Observatory Global Telescope Network, 6740 Cortona Dr., Suite 102, Goleta, CA 93117, USA}
\altaffiltext{10}{Kavli Institute for Theoretical Physics, University of California, Santa Barbara, CA 93106, USA}
\altaffiltext{11}{Inter-University Centre for Astronomy and Astrophysics (IUCAA), Post Bag 4, Ganeshkhind, Pune 411007, India}
\altaffiltext{12}{Jet Propulsion Laboratory, California Institute of Technology, Pasadena, CA 91109, USA}
\altaffiltext{13}{CSPAR and Physics Department, University of Alabama in Huntsville, 320 Sparkman Drive, Huntsville, AL 35899, USA}
\altaffiltext{14}{Texas Tech University, Physics Department, Lubbock, TX 79409-1051, USA}
\altaffiltext{15}{Department of Astronomy and Astrophysics, Pennsylvania State University, University Park, PA 16802, USA}
\altaffiltext{16}{Astrophysics Office, ZP12, NASA Marshall Space Flight Center, Huntsville, AL 35812, USA}
\altaffiltext{17}{Instituto de Astrof\' isica de Andaluc\'ia (IAA-CSIC), Glorieta de la Astronom\'ia s/n, E-18008, Granada, Spain}
\altaffiltext{18}{Unidad Asociada Grupo Ciencia Planetarias UPV/EHU-IAA/CSIC, Departamento de F\'isica Aplicada I, E.T.S. Ingenier\'ia, Universidad del Pa\'is-Vasco UPV/EHU, Alameda de Urquijo s/n, E-48013 Bilbao, Spain}
\altaffiltext{19}{Ikerbasque, Basque Foundation for Science, Alameda de Urquijo 36-5, E-48008 Bilbao, Spain}
\altaffiltext{20}{Benoziyo Center for Astrophysics, Weizmann Institute of Science, 76100 Rehovot, Israel}
\altaffiltext{21}{Space Sciences Laboratory, University of California-Berkeley, Berkeley, CA 94720, USA}
\altaffiltext{22}{The Oskar Klein Centre, Department of Physics, Stockholm University, SE-106 91 Stockholm, Sweden}
\altaffiltext{23}{Th\"uringer Landessternwarte Tautenburg, Sternwarte 5, D-07778 Tautenburg, Germany}
\altaffiltext{24}{Max-Planck Institut f\"ur Extraterrestrische Physik, Giessenbachstrasse 1, D-85748 Garching, Germany}
\altaffiltext{25}{Department of Mathematics and Science, National Taiwan Normal University, Lin-kou District, New Taipei City 24449, Taiwan}
\altaffiltext{26}{Infrared Processing and Analysis Center, California Institute of Technology, Pasadena, CA 91125, USA}
\altaffiltext{27}{Department of Astronomy, University of California, Berkeley, CA 94720-3411, USA}
\altaffiltext{28}{Physics Division, Lawrence Berkeley National Laboratory, Berkeley, CA 94720, USA}
\altaffiltext{29}{Ioffe Physical-Technical Institute, Politekhnicheskaya 26, St Petersburg 194021, Russia}
\altaffiltext{30}{Dark Cosmology Centre, Niels Bohr Institute, Juliane Maries Vej 30, Copenhagen \O, DK-2100, Denmark}
\altaffiltext{31}{Institute of Astronomy, National Central University, Chung-Li 32054, Taiwan}
\altaffiltext{32}{NASA Postdoctoral Fellow}
\altaffiltext{33}{Hubble Fellow}

\shorttitle{The Needle in the 100~\lowercase{deg}$^2$ Haystack}
\shortauthors{Singer et al.}


\begin{abstract}
The \emph{Fermi} Gamma\nobreakdashes-ray Space Telescope has greatly expanded the number and energy window of observations of \acp{GRB}. However, the coarse localizations of tens to a hundred square degrees provided by the \emph{Fermi} GRB Monitor instrument have posed a formidable obstacle to locating the bursts' host galaxies, measuring their redshifts, and tracking their panchromatic afterglows. We have built a target\nobreakdashes-of\nobreakdashes-opportunity mode for the \acl{IPTF} in order to perform targeted searches for \emph{Fermi} afterglows. Here, we present the results of one year of this program: 8 afterglow discoveries out of 35 searches. Two of the bursts with detected afterglows (\acp{GRB}~130702A~and~140606B) were at low redshift ($z = 0.145$ and 0.384 respectively) and had spectroscopically confirmed broad\nobreakdashes-line Type Ic supernovae. We present our broadband follow\nobreakdashes-up including spectroscopy as well as X\nobreakdashes-ray, UV, optical, millimeter, and radio observations. We study possible selection effects in the context of the total \emph{Fermi} and \emph{Swift} \ac{GRB} samples. We identify one new outlier on the Amati relation. We find that two bursts are consistent with a mildly relativistic shock breaking out from the progenitor star, rather than the ultra\nobreakdashes-relativistic internal shock mechanism that powers standard cosmological bursts. Finally, in the context of the \acl{ZTF}, we discuss how we will continue to expand this effort to find optical counterparts of \acl{BNS} mergers that may soon be detected by Advanced \acs{LIGO} and Virgo.
\end{abstract}

\makeatletter{}\section{Introduction}

Deep synoptic optical surveys, including the \acl{PTF} (\acsu{PTF}; \citealt{PTFLaw,PTFRau}) and Pan\nobreakdashes-STARRS \citep{PanSTARRS}, have revealed a wealth of new transient and variable phenomena across a wide range of characteristic luminosities and timescales \citep{KasliwalThesis}. With a wide (7~deg$^2$) instantaneous \ac{FOV}, moderately deep sensitivity (reaching $R = 20.6$\,mag in 60~s), a consortium of follow\nobreakdash-up telescopes, sophisticated image subtraction and machine learning pipelines, and an international team of human\nobreakdashes-in\nobreakdashes-the\nobreakdashes-loop observers, \ac{PTF} has been a wellspring of new or rare kinds of explosive transients (for instance, \citealt{SLSNe,CaRichGapTransients}) and early\nobreakdashes-time observations of \acp{SN} or their progenitors (see, for example, \citealt{PTF11fe,PTF10vgv,PTF10tel,iPTF13ast}). \ac{PTF} has even blindly detected the optical emission (\citealt{GCN15883}; S.~B.~Cenko~et~al., in~preparation) from the rarest, brightest, and briefest of all known cosmic explosions, \acp{GRB}, hitherto only discoverable with the aid of precise localizations from space\nobreakdashes-based gamma\nobreakdashes-ray observatories. \ac{PTF} has also detected explosions that optically resemble \ac{GRB} afterglows but may entirely lack gamma\nobreakdashes-ray emission \citep{PTF11agg}.

\acp{GRB} and their broadband afterglows are notoriously challenging to capture. They naturally evolve from bright to faint, and from high (gamma- and hard X\nobreakdashes-ray) to low (optical and radio) photon energies, with information encoded on energy scales from 1~to~$10^{16}$~GHz \citep{PerleyGRB130427A} and timescales from $10^{-3}$~to~10$^7$~s. Only with a rapid sequence of handoffs between facilities graded by energy passband, \ac{FOV}, and position accuracy have we been able to find them, pinpoint their host galaxies, and constrain their physics. The \emph{Swift} mission \citep{Swift}, with its 1.4~sr\nobreakdashes-wide (50\% coded) \acl{BAT} (\acsu{BAT}; \citealt{BAT}) and its ability to slew and train its onboard \acl{XRT} (\acsu{XRT}; \citealt{XRT}) and \acl{UVOT} (\acsu{UVOT}; \citealt{UVOT}) on the location of a new burst within 100~s, has triumphed here: in nine years of operation, it has tracked down $\approx 700$ X\nobreakdashes-ray afterglows and enabled extensive panchromatic observations by a worldwide collaboration of ground\nobreakdashes-based optical and radio facilities.

Meanwhile, the \emph{Fermi} satellite has opened up a new energy regime extending up to 300~GeV, with the \acl{LAT} (\acsu{LAT}; \citealt{LAT}) detecting high\nobreakdashes-energy photons for about a dozen bursts per year. The \acl{GBM} (\acsu{GBM}; \citealt{GBM}), an all\nobreakdashes-sky instrument sensitive from 8~keV to 40~MeV, detects \acp{GRB} prolifically at a rate of $\approx 250$~yr$^{-1}$, with a large number (about 44~yr$^{-1}$) belonging to the rarer short, hard bursts \citep{FermiGBMCatalog}. Although \ac{LAT} can provide localizations that are as accurate as $\sim$10\arcmin, \emph{Fermi} \ac{GBM} produces error circles that are several degrees across. Since most bursts seem to lack GeV emission detectable by \ac{LAT}, most \emph{Fermi} \ac{GBM} bursts do not receive deep, broadband follow\nobreakdashes-up. Consequently, their redshifts and the properties of their afterglows have remained largely unknown.

As part of the \acf{IPTF}, over the past year we have developed the ability to rapidly tile these $\sim 100$~deg$^2$ \ac{GBM} error circles and pinpoint the afterglows. This \ac{TOO} capability uses and briefly redirects the infrastructure of our ongoing synoptic survey (the operation of which is discussed in \citealt{PTF10vdl}), notably the machine learning software and the instrumental pipeline composed of the \acl{P48} (\acsu{P48}; \citealt{P48PTF}), the \acl{P60} (\acsu{P60}; \citealt{P60Automation}), and associated spectroscopic resources including the \acf{P200}.

In \citet{iPTF13bxl}, we announced the first discovery of an optical afterglow based solely on a \emph{Fermi} \ac{GBM} localization.\footnote{There are two earlier related cases. The optical afterglow of \ac{GRB}\,090902B was detected ex post facto in tiled observations with \ac{ROTSE} about 80~minutes after the burst, but the afterglow was initially discovered with the help of an X\nobreakdashes-ray detection in \emph{Swift} observations of the \ac{LAT} error circle. \ac{GRB}\,120716A was identified by \ac{IPTF} by searching a $\approx 2$~deg$^{2}$ \acs{IPN} error box \citep{GCN13489}.} That explosion, \ac{GRB}\,130702A/iPTF13bxl, was noteworthy for several reasons. First, it was detected by \emph{Fermi} \ac{LAT}. Second, it was at moderately low redshift, $z = 0.145$, yet had prompt energetics that bridged the gap between ``standard,'' bright cosmically distant bursts and nearby sub\nobreakdashes-luminous bursts and \aclp{XRF}. Third, due to its low redshift, an accompanying \ac{SN} was spectroscopically detectable.

In this work, we begin with a detailed description of the operation of the \ac{IPTF} \ac{GRB} afterglow search. We then present seven more \ac{GBM}--\ac{IPTF} afterglows from the first 13 months of this project. In each of the eight cases, the association between the optical transient and the \ac{GRB} was proven by the presence of high\nobreakdashes-redshift absorption lines in the optical spectra and the coincident detection of a rapidly fading X\nobreakdashes-ray source with \emph{Swift} \ac{XRT}. In two cases, the positions were further corroborated by accurate \emph{Fermi} \ac{LAT} error circles, and in four cases by accurate \ac{IPN} triangulations involving distant spacecraft. In one case (\ac{GRB}\,140508A), the \ac{IPN} triangulation was performed rapidly and was instrumental in selecting which optical transient candidates to follow up. In six cases, radio afterglows were detected. Our discovery rate of 8 out of 35 events is consistent with the ages and searched areas of the \ac{GBM} bursts, combined with the luminosity function of optical afterglows. Consequently, by tiling larger areas and/or stacking exposures, the \ac{IPTF} afterglow search should be able to scale to coarser localizations and fainter optical signals, such as those associated with short \acp{GRB}.

Next, we present extensive follow\nobreakdashes-up observations, including $R$\nobreakdashes-band photometry from the \ac{P48}, multicolor photometry from the \ac{P60}, spectroscopy (acquired with the \ac{P200}, Keck, Gemini, APO, \emph{Magellan}, \ac{VLT}, and GTC), and radio observations with the \acl{VLA}\footnote{\url{http://www.vla.nrao.edu}} (\acsu{VLA}), the \acl{CARMA} (\acsu{CARMA}; \citealt{CARMA1,CARMA2}), the \acl{ATCA} (\acsu{ATCA}; \citealt{ATCA}), and the \acl{AMI} (\acsu{AMI}; \citealt{AMI}). We provide basic physical interpretations of the broadband \acp{SED} of these afterglows. We find that seven of the events are consistent with the classic model of synchrotron cooling of electrons that have been accelerated by a single forward shock encountering either the constant-density circumburst \acl{ISM} (\acsu{ISM}; broadband behavior predicted in \citealt{AfterglowSpectra}) or a stellar (i.e., Wolf\nobreakdashes--Rayet) wind environment \citep{AfterglowSpectraWind}. The possible exception, \ac{GRB}\,140620A/iPTF14cva, can probably be explained by standard extensions of this model, a reverse shock or an inverse Compton component.

Two of the afterglows (\ac{GRB}\,130702A/iPTF13bxl and \ac{GRB}\,140606B/iPTF14bfu) faded away to reveal spectroscopically detected \acp{SNIcBL}. Despite the abundant photometric evidence for \acp{SN} in afterglow light curves (see \citealt{LightCurvesGRBSNe} and references therein), the distinction of \ac{SN} spectroscopy has been shared by scarcely tens\footnote{Between photometric, late\nobreakdashes-time red bumps and unambiguous spectral identifications, there are also \ac{GRB}\nobreakdashes--\acp{SN} that have some \ac{SN}\nobreakdashes-associated spectral features. The number of \acp{GRB} with spectroscopic \acp{SN} is, therefore, ill defined. See \citet[p. 169, and references therein]{HjorthGRBSNConnection} for a more complete census.} out of $\approx$800 long \emph{Swift} bursts in nine years of operation.

We estimate the kinetic energies of the relativistic blast waves of these events from their X\nobreakdashes-ray afterglows \citep{EnergyOfGammaRayBursts}. We find that although the gamma\nobreakdashes-ray energetics of these eight bursts are broadly similar to the \emph{Swift} sample, two low\nobreakdashes-luminosity bursts (\acp{GRB}\,130702A~and~140606B) have significantly lower kinetic energies. We discuss the possibility that these two bursts arise not from a standard ultra\nobreakdashes-relativistic internal shock, but from a mildly relativistic shock as it breaks out from the progenitor star (see, for example, \citealt{RelativisticShockBreakoutRelation}).

We conclude by discussing prospects for targeted optical transient searches in wide areas. This is especially relevant for optical counterparts of \ac{GW} events. We illustrate that optical afterglows of short bursts, which are intimately linked to the prime sources for the Advanced \acf{LIGO} and Virgo, should be well within the reach of a similar approach using \acl{ZTF} (\acsu{ZTF}; \citealt{ZTF,ZTFBellm,ZTFSmith}).

\section{Search Methodology}
\label{sec:afterglow-search-method}

We begin by describing our \ac{TOO} observations and afterglow search step by step.

\subsection{Automated \acs{TOO} Marshal: Alerts and Tiling}

A program called the \ac{IPTF} \ac{TOO} Marshal monitors the stream of \ac{GCN} notices\footnote{\url{http://gcn.gsfc.nasa.gov}} from the three redundant, anonymous NASA/GSFC VOEvent servers. It listens for notices of type \texttt{FERMI\_GBM\_GND\_POS}, sent by \ac{GBM}'s automated on\nobreakdashes-ground localization, or \texttt{FERMI\_GBM\_FIN\_POS}, sent by the \ac{GBM} burst advocate's human\nobreakdashes-in\nobreakdashes-the\nobreakdashes-loop localization.\footnote{Usually, the \emph{Fermi} team suppresses the notices if the burst is detected and localized more accurately by \emph{Swift} \ac{BAT}.}

Upon receiving either kind of notice, the \ac{TOO} Marshal determines if the best-estimate sky position is observable from Palomar at any time within the 24 hr after the trigger. The criterion for observability is that the position is at an altitude $> 23\fdg5$ (i.e. airmass $\lesssim 2.5$), at least $20\arcdeg$ from the center of the moon, at an hour angle between $\pm 6\fh5$, and that the Sun is at least $12\arcdeg$ below the horizon at Palomar.

If the position is observable and the 1$\sigma$ statistical error radius $r_\mathrm{stat}$ reported in the \ac{GCN} notice is less than $10\arcdeg$, the \ac{TOO} Marshal selects a set of 10 \ac{P48} fields that optimally cover the error region.\footnote{We made one exception to our \ac{GBM} error radius cutoff: we followed up \ac{GRB}\,140219A, which had a \ac{GBM} error circle with a radius of $12.8\arcdeg$, but had an \ac{IPN} localization spanning 0.6~deg$^2$ \citep{GCN15864}. Despite searching about 80\% of the \ac{IPN} polygon, we detected no afterglow \citep{GCN15878}. This is potentially a dark burst candidate.} It converts the \ac{GBM} position estimate and radius into a probability distribution by applying a well\nobreakdashes-known empirical prescription of the systematic errors of the \ac{GBM} localization. \citet{FermiGBMFirstTwoYears} state that the total effective error radius in the \texttt{FERMI\_GBM\_FIN\_POS} localizations is well described by the quadrature sum of the statistical radius and a systematic contribution, where the systematic is $2\fdg6$ for 72\% of bursts and $10\fdg4$ for 28\% of bursts. We use the weighted rms of these two values, $r_\mathrm{sys} = \sqrt{0.72(2\fdg6)^2 + 0.28(10\fdg4)^2} \approx 6\arcdeg$. The total error radius is then $r_\mathrm{eff} = \sqrt{{r_\mathrm{stat}}^2 + {r_\mathrm{sys}}^2}$. We construct a Fisher\nobreakdashes--von~Mises distribution, centered on the best\nobreakdashes-estimate position, with a concentration parameter of
\begin{equation}
    \kappa = \left[1 - \cos \left( \frac{\pi}{180\arcdeg} r_\mathrm{eff} \right)\right]^{-1}.
\end{equation}

With the \texttt{FERMI\_GBM\_FIN\_POS} alert, the \emph{Fermi} \ac{GBM} team also distributes a detailed localization map that accounts for the systematic effects \citep{GBMLocalization}. The \ac{TOO} Marshal retrieves from the \emph{Fermi} data archive a file that describes the 1$\sigma$, 2$\sigma$, and 3$\sigma$ significance contours. If the localization has significant asymmetry, we also retrieve a 2D FITS image whose pixel values correspond to the \ac{GBM} localization significance, and use this instead of the Fisher\nobreakdashes--von~Mises distribution.

Giving preference to fields for which deep co\nobreakdashes-added reference images exist, the \ac{TOO} Marshal selects 10 \ac{P48} fields spanning an area of $\approx 72$~deg$^2$ to maximize the probability of enclosing the true (but as yet unknown) location of the source, assuming the above distribution.

The Marshal then immediately contacts a team of humans (the authors) by SMS text message, telephone, and e-mail. The humans are directed to a mobile\nobreakdashes-optimized web application to trigger the \ac{P48} (see Fig.~\ref{fig:screenshot}).

\subsection{Triggering the \acs{P48}}

Within the above constraints, we decide whether to follow up the burst based on the following criteria. The event must be $\lesssim$12\,hr old when it first becomes observable from Palomar, and we must cover enough of the error circle to have a $\gtrsim$30\% chance of enclosing the position of the source. We discard any bursts that are detected and accurately localized by \emph{Swift} \ac{BAT}, because these are more efficiently followed up by conventional means. We also give preference to events that are out of the Galactic plane and that are observable for at least 3\,hr.

There are some exceptional circumstances that override these considerations. If the burst's position estimate is accessible within an hour after the burst, we may select it even if the observability window is very brief. If the burst is very well localized or has the possibility of a substantially improved localization later due to a \ac{LAT} or \ac{IPN} detection, we may select it even if it is in the Galactic plane.

The default observing program is three epochs of \ac{P48} images at a 30\nobreakdashes-minute cadence. The human may shorten or lengthen the cadence if the burst is very young or old (see the discussion of Equation~(\ref{eq:subtraction-sensitivity-loss}) in Section~\ref{sec:visual-scanning-in-treasures-portal} below), change the number of epochs, or add and remove \ac{P48} fields. When the human presses the ``Go'' button, the \ac{TOO} Marshal sends a machine-readable e-mail to the \ac{P48} robot. The robot adds the requested fields to the night's schedule with the highest possible priority, ensuring that they are observed as soon as they are visible.

\subsection{Automated Candidate Selection}

As the night progresses, the \ac{TOO} Marshal monitors the progress of the observations and the \ac{IPTF} real\nobreakdashes-time image subtraction pipeline (P.~E.~Nugent~et~al. 2015, in~preparation). The real\nobreakdashes-time pipeline creates difference images between the new \ac{P48} observations and co\nobreakdashes-added references composed of observations from months or years earlier. It generates candidates by performing source extraction on the difference images. A machine learning classifier assigns a \emph{real/bogus} score (RB2; \citealt{RB2}) to each candidate that predicts how likely the candidate is to be a genuine astrophysical source (rather than a radiation hit, a ghost, an imperfect image subtraction residual, or any other kind of artifact).

Table~\ref{table:vetting} lists the number of candidates that remain after each stage of candidate selection. First, requiring candidates to have \ac{SNR}$>5$ gives us a median of 35,000 candidates. This number varies widely with galactic latitude and the area searched (a median of $\sim$500~deg$^{-2}$). Second, we only select candidates that have RB2$>0.1$, reducing the number of candidates to a median of 36\% of the original list.\footnote{This RB2 threshold is somewhat deeper than that which is used in the \ac{IPTF} survey. An improved classifier, RB4 \citep{RB4}, entered evaluation in 2014 August shortly before \ac{GRB}\,140808A.} Third, we reject candidates that coincide with known stars in reference catalogs (\ac{SDSS} and the \ac{PTF} reference catalog), cutting the list to 17\%. Fourth, we eliminate asteroids cataloged by the Minor Planet Center, reducing the list to 16\%. Fifth, we demand at least two secure \ac{P48} detections after the \ac{GBM} trigger, reducing the list to a few percent, or $\sim 500$ candidates.

When the image subtraction pipeline has finished analyzing at least two successive epochs of any one field, the \ac{TOO} Marshal contacts the humans again and the surviving candidates are presented to the humans via the Treasures portal.

\begin{deluxetable*}{rrrrrrrr}
\tablewidth{\textwidth}
\tablecaption{\label{table:vetting}Number of Optical Transient Candidates Surviving Each Vetting Stage}
\tablehead{
    \colhead{} &
    \colhead{\acs{SNR}} &
    \colhead{RB2} &
    \colhead{Not} &
    \colhead{Not in} &
    \colhead{Detected} &
    \colhead{Saved for} &
    \colhead{} \\
    \colhead{GRB} &
    \colhead{$>5$} &
    \colhead{$>0.1$} &
    \colhead{Stellar} &
    \colhead{MPC\tablenotemark{a}} &
    \colhead{Twice} &
    \colhead{Follow\nobreakdashes-up} &
    \colhead{RB2\tablenotemark{b}}
}
\startdata
130702A  &   14\,629  &   2\,388  &   1\,346  &   1\,323  &     417  &  11 & 0.843 \\
131011A  &   21\,308  &   8\,652  &   4\,344  &   4\,197  &     434  &  23 & 0.198 \\
131231A  &    9\,843  &   2\,503  &   1\,776  &   1\,543  &  1\,265  &  10 & 0.137 \\
140508A  &   48\,747  &  22\,673  &   9\,970  &   9\,969  &     619  &  42 & 0.730 \\
140606B  &   68\,628  &  26\,070  &  11\,063  &  11\,063  &  1\,449  &  28 & 0.804 \\
140620A  &  152\,224  &  50\,930  &  17\,872  &  17\,872  &  1\,904  &  34 & 0.826 \\
140623A  &   71\,219  &  29\,434  &  26\,279  &  26\,279  &     442  &  23 & 0.873 \\
140808A  &   19\,853  &   4\,804  &   2\,349  &   2\,349  &      79  &  12 & 0.318 \\
\tableline
\multicolumn{2}{r}{Median reduction} & 36\% & 17\% & 16\% & 1.7\% & 0.068\% &
\enddata
\tablenotetext{a}{Not in Minor Planet Center database}
\tablenotetext{b}{RB2 score of optical afterglow in earliest \acs{P48} detection}
\end{deluxetable*}

\subsection{Visual scanning in Treasures Portal}
\label{sec:visual-scanning-in-treasures-portal}

The remaining candidate vetting steps currently involve human participation and are informed by the nature of the other transients that \ac{IPTF} commonly detects: foreground \acp{SN} (slowly varying and in low\nobreakdashes-$z$ host galaxies), \acp{AGN}, cataclysmic variables, and M\nobreakdashes-dwarf flares.

In the Treasures portal, we visually scan through the automatically selected candidates one \ac{P48} field at a time, examining $\sim$10 objects per field (see Figure~\ref{fig:treasures} for a screenshot of the Treasures portal). We visually assess each candidate's image subtraction residual compared to the neighboring stars of similar brightness in the new image. If the residual resembles the new image's \acl{PSF}, then the candidate is considered likely to be a genuine transient or variable source.

Next, we look at the photometric history of the candidates. Given the time, $t$, of the optical observation relative to the burst and the cadence, $\delta t$, we expect that a typical optical afterglow that decays as a power law $F_\nu \propto t^{-\alpha}$, with $\alpha=1$, would fade by $\delta m = 2.5 \log_{10} (1+\delta t/t)$~mag over the course of our observations. Any source that exhibits statistically significant fading ($\delta m / m \gg 1$) consistent with an afterglow decay becomes a prime target.\footnote{A source that exhibits a statistically significant rise is generally also followed up, but as part of the main \ac{IPTF} transient survey, rather than as a potential optical afterglow.}

Note that a $1\sigma$ decay in brightness requires such a source to be
\begin{equation}\label{eq:subtraction-sensitivity-loss}
    -2.5\log_{10} \left(\frac{\delta t}{t\sqrt{2}}\right)
\end{equation}
brighter than the $1\sigma$ limiting magnitude of the exposures. For example, given the \ac{P48}'s typical limiting magnitude of $R = 20.6$ and the standard cadence of $\delta t = 0.5$\,hr, if a burst is observed $t = 3$\,hr after the trigger, its afterglow may be expected to have detectable photometric evolution only if it is brighter than $R = 18.3$. Noting that long \acp{GRB} preferentially occur at high redshifts and in intrinsically small, faint galaxies \citep{GRBSNHostGalaxies}, we consider faint sources that do not display evidence of fading if they are not spatially coincident with any sources in \ac{SDSS} or archival \ac{IPTF} observations.

If a faint source is near a spatially resolved galaxy, then we compute its distance modulus using the galaxy's redshift or photometric redshift from SDSS. We know that long \ac{GRB} optical afterglows at $t=1$~day typically have absolute magnitudes of $-25 < M_B < -21$\,mag (1$\sigma$ range; see Figure~9 of \citealt{KannTypeITypeIIOpticalAfterglows}). Most \acp{SN} are significantly fainter: Type~Ia are typically $M_B \sim -19$\,mag whereas Ibc and II are $M_B \sim -17$\,mag, with luminous varieties of both Type~Ibc and II extending to $M_B \sim -19$\,mag \citep{RichardsonComparativeSupernovae,LickSupernovaLuminosityFunction}. Therefore, if the candidate's presumed host galaxy would give it an absolute magnitude $M_R < -20$~mag, it is considered promising. This criterion is only useful for long \acp{GRB} because short \ac{GRB} afterglows are typically $\sim 6$~mag fainter than long \ac{GRB} afterglows \citep{KannTypeITypeIIOpticalAfterglows}.

The human saves all candidates that are considered promising by these measures to the \ac{IPTF} Transient Marshal database. This step baptizes them with an \ac{IPTF} transient name, which consists of the last two digits of the year and a sequential alphabetic designation.

\subsection{Archival vetting in the Transient Marshal}

Once named in the Transient Marshal, we perform archival vetting of each candidate using databases including VizieR \citep{VizieR}, NED,\footnote{http://ned.ipac.caltech.edu} the \acf{HEASARC},\footnote{http://heasarc.gsfc.nasa.gov} and \acl{CRTS} \citep{CRTS}, in order to check for any past history of variability at that position (see Figure~\ref{fig:transient-marshal} for a screenshot of the Transient Marshal).

We check for associations with known quasars or \acp{AGN} in \citet{QuasarAtlas} or with \ac{AGN} candidates in \citet{ARXA}.

M~dwarfs can produce bright, blue, rapidly fading optical flares than can mimic optical afterglows. To filter our M~dwarfs, we check for quiescent infrared counterparts in \emph{WISE} \citep{ALLWISE}. Stars of spectral type L9\nobreakdashes--M0 peak slightly blueward of the \emph{WISE} bandpass, with typical colors \citep{WISEOnOrbit}
\begin{eqnarray*}
    3 \lesssim& [R-W1] &\lesssim 12 \\
    0.1 \lesssim& [W1-W2] &\lesssim 0.6 \\
    0.2 \lesssim& [W2-W3] &\lesssim 1 \\
    0 \lesssim& [W3-W4] &\lesssim 0.2.
\end{eqnarray*}
Therefore, a source that is detectable in \emph{WISE} but that is either absent from or very faint in the \ac{IPTF} reference images suggests a quiescent dwarf star.

\subsection{Photometric, Spectroscopic, and Broad-band Follow-up}

The above stages usually result in $\sim$10 promising optical transient candidates that merit further follow\nobreakdashes-up. If, by this point, data from \emph{Fermi} \ac{LAT} or from \ac{IPN} satellites are available, we can use the improved localization to select an even smaller number of follow\nobreakdashes-up targets.

For sources whose photometric evolution is not clear, we perform photometric follow\nobreakdashes-up. We may schedule additional observations of some of the \ac{P48} fields if a significant number of candidates are in the same field. We may also use the \ac{P48} to gather more photometry for sources that are superimposed on a quiescent source or galaxy, in order to make use of the image subtraction pipeline to automatically obtain host\nobreakdashes-subtracted magnitudes. For isolated sources, we schedule one or more epochs of $r$\nobreakdashes-band photometry with the \ac{P60}. If, by this point, any candidates show strong evidence of fading, we begin multicolor photometric monitoring with the \ac{P60}.

Next, we acquire spectra for one to three candidates per burst using the \ac{P200}, Gemini, Keck, Magellan, or \ac{HCT}. A spectrum that has a relatively featureless continuum and high\nobreakdashes-redshift absorption lines secures the classification of the candidate as an optical afterglow.

Once any single candidate becomes strongly favored over the others based on photometry or spectroscopy, we trigger X\nobreakdashes-ray and UV observations with \emph{Swift} and radio observations with \ac{VLA}, \ac{CARMA}, and \ac{AMI}. Detection of a radio or X\nobreakdashes-ray afterglow typically confirms the nature of the optical transient, even without spectroscopy.

Finally, we promptly release our candidates, upper limits, and/or confirmed afterglow discovery in \ac{GCN} circulars.

\subsection{Long-term Monitoring and Data Reduction}

The reported \ac{P48} magnitudes are all in the Mould $R$ band and in the AB system \citep{ABMags}, calibrated with respect to either $r^{\prime}$ point sources from \ac{SDSS} or for non\nobreakdashes-\ac{SDSS} fields using the methods described in \citet{PTFPhotometricCalibration}.

To monitor the optical evolution of afterglows identified by our program, we typically request nightly observations in $ri$ (and occasionally $gz$) filters for as long as the afterglow remained detectable. Bias subtraction, flat-fielding, and other basic reductions are performed automatically at Palomar by the \ac{P60} automated pipeline using standard techniques. Images are then downloaded and stacked as necessary to improve the \ac{SNR}. Photometry of the optical afterglow is then performed in IDL using a custom aperture\nobreakdashes-photometry routine, calibrated relative to \ac{SDSS} secondary standards in the field (when available) or using our own solution for secondary field standards constructed during a photometric night (for fields outside the \ac{SDSS} footprint).

For some bursts (GRB~140606B), we also obtain photometry with the \acf{LMI} mounted on the 4.3\,m \acf{DCT} in Happy Jack, Arizona. Standard CCD reduction techniques (e.g., bias subtraction, flat\nobreakdashes-fielding) are applied using a custom IRAF pipeline. Individual exposures are aligned with respect to astrometry from the \acl{2MASS} (\acsu{2MASS}; \citealt{2MASS}) using SCAMP \citep{SCAMP} and stacked with SWarp \citep{SWarp}.

Where GROND \citep{GROND} and RATIR \citep{RATIR} have reported multicolor photometry in \ac{GCN} circulars, we include their published data in Table~\ref{tab:photometry} and our light\nobreakdashes-curve plots.

We monitor \ac{GBM}--\ac{IPTF} afterglows with \ac{CARMA}, a millimeter\nobreakdashes-wave interferometer located at Cedar Flat near Big Pine, California. All observations are conducted at 93~GHz in single-polarization mode in the array's C, D, or E configuration. Targets are typically observed once for 1\nobreakdashes--3\,hr within a few days after the \ac{GRB}, establishing the phase calibration using periodic observations of a nearby phase calibrator and the bandpass and the flux calibration by observations of a standard source at the start of the track. If detected, we acquire additional observations in approximately logarithmically spaced time intervals until the afterglow flux falls below detection limits. All observations are reduced using MIRIAD using standard interferometric flagging and cleaning procedures.

We look for radio afterglows at 6.1 and/or 22~GHz with \ac{VLA}. \ac{VLA} observations are reduced using the \ac{CASA} package. The calibration is performed using the \ac{VLA} calibration pipeline. After running the pipeline, we inspect the data (calibrators and target source) and apply further flagging when needed. The \ac{VLA} measurement errors are a combination of the rms map error, which measures the contribution of small unresolved fluctuations in the background emission and random map fluctuations due to receiver noise, and a basic fractional error (here estimated to be $\approx 5\%$) which accounts for inaccuracies of the flux density calibration. These errors are added in quadrature, and total errors are reported in Table~\ref{tab:radio}.

Starting in 2014 August, we also look for radio emission with \ac{AMI}. \ac{AMI} is composed of eight 12.8~m dishes operating in the 13.9\nobreakdashes--17.5~GHz range (central frequency of 15.7~GHz) when using frequency channels 3\nobreakdashes--7 (channels 1, 2, and 8 are disregarded due to their currently susceptibility to radio interference). For further details on the reduction and analysis performed on the AMI observations please see \citet{GRB130427A-AMI}.

\section{The \acs{GBM}--\acs{IPTF} bursts}

\begin{figure*}
    \centering
    \includegraphics[width=\textwidth]{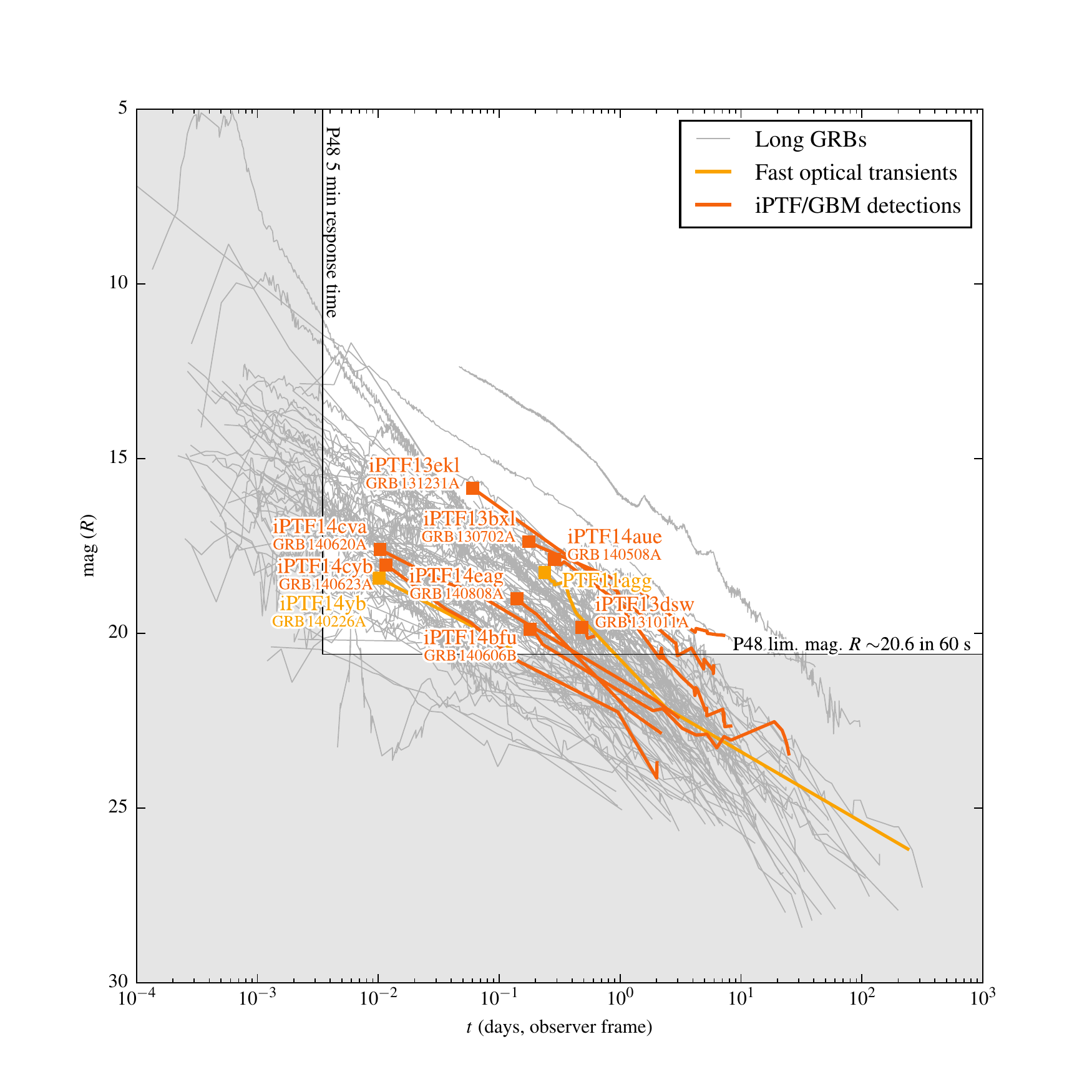}
    \caption[Optical light curves of long \acsp{GRB}]{\label{fig:lightcurve-zoo}Optical light curves of \emph{Fermi}\nobreakdashes--\ac{IPTF} afterglows to date. The light curves of the eight \ac{IPTF}/\ac{GBM} bursts are shown in red. For comparison, the gray lines show a comprehensive sample of long \ac{GRB} optical light curves from \citet{DarkBurstsSwiftEra}, \citet{KanSwiftAfterglowsI}, \citet{PerleyGRB130427A}, and D. A. Kann (2015, private communication). The white area outside of the light\nobreakdashes-gray shading illustrates the range of \ac{GRB} afterglows that are accessible given a half\nobreakdashes-hour cadence and the \ac{P48}'s 60~s limiting magnitude of $R = 20.6$. The two light curves shown in blue are other related \ac{IPTF} transients. The first is PTF11agg, an afterglow\nobreakdashes-like transient with no detected gamma\nobreakdashes-ray emission \citep{PTF11agg}. The second is \ac{GRB}\,140226A/iPTF14yb, reported initially by \ac{IPTF} from its optical afterglow (\citealt{GCN15883}; Cenko et al., in preparation), and later by \ac{IPN} from its gamma-ray emission \citep{GCN15888}.}
\end{figure*}

To date, we have successfully followed up 35 \emph{Fermi} \ac{GBM} bursts and detected eight optical afterglows. The detections are listed in Table~\ref{table:detections}, and all of the \ac{P48} tilings are listed in  Table~\ref{table:nondetections}. Figure~\ref{fig:discovery-images} shows the \ac{GBM} localizations and \ac{P48} tilings for the detected bursts. In Figure~\ref{fig:lightcurve-zoo}, the light curves are shown in the context of a comprehensive sample of long \ac{GRB} afterglows compiled by D.~A.~Kann~(2015, private communication).

{
\makeatletter{}
\begin{deluxetable*}{ll|rrr|lr@{$\pm$}lr@{$\pm$}lr@{$\pm$}lr}
\tablewidth{\textwidth}
\tablecaption{\label{table:detections}\acs{GBM}--\acs{IPTF} Detections}
\tablehead{
    \colhead{} & \colhead{} & \colhead{R.A.} & \colhead{Decl.} & \colhead{Gal.} & \colhead{} & \multicolumn{2}{c}{$E_\mathrm{peak}$} & \multicolumn{2}{c}{$E_{\gamma,\mathrm{iso}}$} & \colhead{} & \colhead{} & \colhead{} \\
    \colhead{GRB} & \colhead{OT} & \colhead{(J2000)} & \colhead{(J2000)} & \colhead{Lat.\tablenotemark{a}} & \colhead{$z$} & \multicolumn{2}{c}{(keV, rest)} & \multicolumn{2}{c}{($10^{52}$\,erg, rest)\tablenotemark{b,c}} & \multicolumn{2}{c}{$T_{90}$ (s)} & \colhead{$m_R(t_\mathrm{P48})\tablenotemark{d}$}
}
\startdata
    GRB\,130702A & iPTF13bxl & 14$^\mathrm{h}$29$^\mathrm{m}$15$^\mathrm{s}$ & +15$\arcdeg$46$\arcmin$26$\arcsec$ & 65$\arcdeg$ & 0.145 & 18 & 3 & $<$0.065 & 0.001 & 58.9 & 6.2 & 17.38 \\
    GRB\,131011A & iPTF13dsw & 02$^\mathrm{h}$10$^\mathrm{m}$06$^\mathrm{s}$ & -4$\arcdeg$24$\arcmin$40$\arcsec$ & -61$\arcdeg$ & 1.874 & 625 & 92 & 14.606 & 1.256 & 77.1 & 3 & 19.83 \\
    GRB\,131231A & iPTF13ekl & 00$^\mathrm{h}$42$^\mathrm{m}$22$^\mathrm{s}$ & -1$\arcdeg$39$\arcmin$11$\arcsec$ & -64$\arcdeg$ & 0.6419 & 291 & 6 & 23.015 & 0.278 & 31.2 & 0.6 & 15.85 \\
    GRB\,140508A & iPTF14aue & 17$^\mathrm{h}$01$^\mathrm{m}$52$^\mathrm{s}$ & +46$\arcdeg$46$\arcmin$50$\arcsec$ & 38$\arcdeg$ & 1.03 & 534 & 28 & 24.529 & 0.86 & 44.3 & 0.2 & 17.89 \\
    GRB\,140606B & iPTF14bfu & 21$^\mathrm{h}$52$^\mathrm{m}$30$^\mathrm{s}$ & +32$\arcdeg$00$\arcmin$51$\arcsec$ & -17$\arcdeg$ & 0.384 & 801 & 182 & 0.468 & 0.04 & 22.8 & 2.1 & 19.89 \\
    GRB\,140620A & iPTF14cva & 18$^\mathrm{h}$47$^\mathrm{m}$29$^\mathrm{s}$ & +49$\arcdeg$43$\arcmin$52$\arcsec$ & 21$\arcdeg$ & 2.04 & 387 & 34 & 7.28\phn & 0.372 & 45.8 & 12.1 & 17.60 \\
    GRB\,140623A & iPTF14cyb & 15$^\mathrm{h}$01$^\mathrm{m}$53$^\mathrm{s}$ & +81$\arcdeg$11$\arcmin$29$\arcsec$ & 34$\arcdeg$ & 1.92 & 834 & 317 & 3.58\phn & 0.398 & 114.7 & 9.2 & 18.04 \\
    GRB\,140808A & iPTF14eag & 14$^\mathrm{h}$44$^\mathrm{m}$53$^\mathrm{s}$ & +49$\arcdeg$12$\arcmin$51$\arcsec$ & 59$\arcdeg$ & 3.29 & 503 & 35 & 8.714 & 0.596 & 4.5 & 0.4 & 19.01
\enddata
\tablenotetext{a}{Galactic latitude of optical afterglow. This is one of the main factors that influences the number of optical transient candidates in Table~\ref{table:vetting}.}
\tablenotetext{b}{$E_{\gamma,\mathrm{iso}}$ is given for a 1\,keV\nobreakdashes--10\,MeV rest\nobreakdashes-frame bandpass.}
\tablenotetext{c}{The rest\nobreakdashes-frame spectral properties, $E_\mathrm{peak}$ and $E_{\gamma,\mathrm{iso}}$, for GRB~130702A are reproduced from \citet{GCN15025}. For all other bursts, we calculated these quantities from the spectral fits (the \texttt{scat} files) in the \emph{Fermi} \ac{GBM} catalog \citep{GBMSpectralCatalog} using the $k$\nobreakdashes-correction procedure described by \citet{KCorrectionGRBs}.}
\tablenotetext{d}{$R$\nobreakdashes-band apparent magnitude in initial \ac{P48} detection.}
\end{deluxetable*}
 
}

{
\makeatletter{}
\begin{deluxetable}{r@{}lr@{$\pm$}lrrrr}
\tablewidth{\columnwidth}
\tablecaption{\label{table:nondetections}Log of \acs{P48} Tilings for \emph{Fermi} \acs{GBM} Bursts}
\tablehead{
    \colhead{} & \colhead{} & \multicolumn{2}{c}{\acs{GBM}} & \colhead{$t_\mathrm{P48}$} & \colhead{P48} & \colhead{} & \colhead{} \\
    \colhead{} & \colhead{GRB Time\tablenotemark{a}} & \multicolumn{2}{c}{Fluence\tablenotemark{b}} & \colhead{$-t_\mathrm{burst}$\tablenotemark{c}} & \colhead{Area\tablenotemark{d}} & \colhead{Prob.\tablenotemark{e}} & \colhead{}
}
\startdata
     &
    2013 Jun 28 20:37:57 & 10\phd\phn & 0.1 & 10.02 & 73 & 32\% \\
    $\rightarrow$\bfseries &
    2013 Jul 02 00:05:20 & 57\phd\phn & 1.2 & 4.20 & 74 & 38\% \\
     &
    2013 Aug 28 07:19:56 & 372\phd\phn & 0.6 & 20.28 & 74 & 64\% \\
     &
    2013 Sep 24 06:06:45 & 37\phd\phn & 0.6 & 23.24 & 74 & 28\% \\
     &
    2013 Oct 06 20:09:48 & 18\phd\phn & 0.6 & 15.26 & 74 & 18\% \\
    $\rightarrow$\bfseries &
    2013 Oct 11 17:47:30 & 89\phd\phn & 0.6 & 11.56 & 73 & 54\% \\
     &
    2013 Nov 08 00:34:39 & 28\phd\phn & 0.5 & 4.69 & 73 & 37\% \\
     &
    2013 Nov 10 08:56:58 & 33\phd\phn & 0.3 & 17.47 & 73 & 44\% \\
     &
    2013 Nov 25 16:32:47 & 5.5 & 0.3 & 11.72 & 95 & 26\% \\
     &
    2013 Nov 26 03:54:06 & 17\phd\phn & 0.3 & 6.94 & 109 & 59\% \\
     &
    2013 Nov 27 14:12:14 & 385\phd\phn & 1.4 & 13.46 & 60 & 50\% \\
     &
    2013 Dec 30 19:24:06 & 41\phd\phn & 0.4 & 7.22 & 80 & 38\% \\
    $\rightarrow$\bfseries &
    2013 Dec 31 04:45:12 & 1519\phd\phn & 1.2 & 1.37 & 30 & 32\% \\
     &
    2014 Jan 04 17:32:00 & 333\phd\phn & 0.6 & 18.57 & 15 & 11\% \\
     &
    2014 Jan 05 01:32:57 & 6.4 & 0.1 & 7.63 & 74 & 22\% \\
     &
    2014 Jan 22 14:19:44 & 9.1 & 0.5 & 11.97 & 75 & 34\% \\
     &
    2014 Feb 11 02:10:41 & 7.4 & 0.3 & 1.77 & 44 & 19\% \\
     &
    2014 Feb 19 19:46:32 & 28\phd\phn & 0.5 & 7.01 & 71 & 14\% \\
     &
    2014 Feb 24 18:55:20 & 24\phd\phn & 0.6 & 7.90 & 72 & 30\% \\
     &
    2014 Mar 11 14:49:13 & 40\phd\phn & 1.2 & 12.18 & 73 & 54\% \\
     &
    2014 Mar 19 23:08:30 & 71\phd\phn & 0.3 & 3.88 & 74 & 48\% \\
     &
    2014 Apr 04 04:06:48 & 82\phd\phn & 0.2 & 0.11 & 109 & 69\% \\
     &
    2014 Apr 29 23:24:42 & 6.2 & 0.2 & 10.99 & 74 & 15\% \\
    $\rightarrow$\bfseries &
    2014 May 08 03:03:55 & 614\phd\phn & 1.2 & 6.68 & 73 & 67\% \\
     &
    2014 May 17 19:31:18 & 45\phd\phn & 0.4 & 8.60 & 95 & 69\% \\
     &
    2014 May 19 01:01:45 & 39\phd\phn & 0.5 & 4.42 & 73 & 41\% \\
    $\rightarrow$\bfseries &
    2014 Jun 06 03:11:52 & 76\phd\phn & 0.4 & 4.08 & 74 & 56\% \\
     &
    2014 Jun 08 17:07:11 & 19\phd\phn & 0.6 & 11.20 & 73 & 49\% \\
    $\rightarrow$\bfseries &
    2014 Jun 20 05:15:28 & 61\phd\phn & 0.6 & 0.17 & 147 & 59\% \\
    $\rightarrow$\bfseries &
    2014 Jun 23 05:22:07 & 61\phd\phn & 0.6 & 0.18 & 74 & 4\% \\
     &
    2014 Jun 28 16:53:19 & 18\phd\phn & 1.0 & 16.16 & 76 & 20\% \\
     &
    2014 Jul 16 07:20:13 & 2.4 & 0.3 & 0.17 & 74 & 28\% \\
     &
    2014 Jul 29 00:36:54 & 81\phd\phn & 0.7 & 3.43 & 73 & 65\% \\
     &
    2014 Aug 07 11:59:33 & 13\phd\phn & 0.1 & 15.88 & 73 & 54\% \\
    $\rightarrow$\bfseries &
    2014 Aug 08 00:54:01 & 32\phd\phn & 0.3 & 3.25 & 95 & 69\%
\enddata
\tablenotetext{a}{Time of \emph{Fermi} \ac{GBM} trigger. $\rightarrow$Afterglow detections are marked with an arrow. The corresponding entries in Table~\ref{table:detections} can be found by matching the date to the \ac{GRB} name (\ac{GRB}~YYMMDDA).}
\tablenotetext{b}{Observed \emph{Fermi} \acs{GBM} fluence in the 10\nobreakdashes--1000~keV band, in units of~$10^{-7}$~erg\,cm$^{-2}$. This quantity is taken from the \texttt{bcat} files from the \emph{Fermi} \ac{GRB} catalog at \ac{HEASARC}.}
\tablenotetext{c}{Age in hours of the burst at the beginning of the \ac{P48} observations.}
\tablenotetext{d}{Area in deg$^2$ spanned by the \ac{P48} fields.}
\tablenotetext{e}{Probability, given the \emph{Fermi} \ac{GBM} localization, that the source is contained within the \ac{P48} fields.}
\end{deluxetable}
 
}

\begin{figure}
    \centering
    \includegraphics{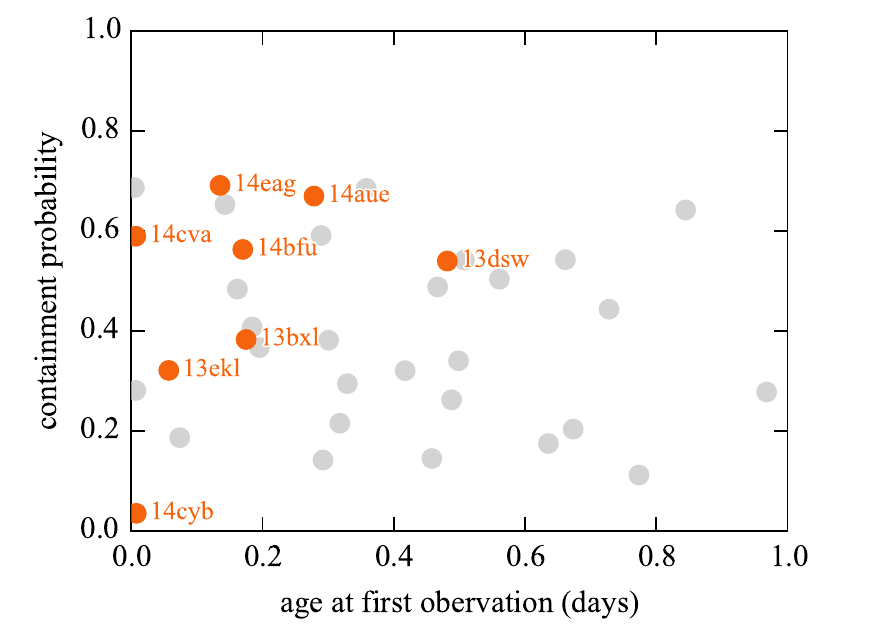}
    \caption[Containment probability and age]{\label{fig:time-prob}
    Prior probability of containing the burst's location within the \ac{P48} fields vs. age of the burst at the beginning of \ac{P48} observations. Afterglow detections are shown in orange, and non\nobreakdashes-detections are shown in gray.}
\end{figure}

The outcome of an individual afterglow search is largely determined by two factors: how much probability is contained within the \ac{P48} footprints, and how bright the afterglow is at the time of the observations (see Figure~\ref{fig:time-prob}). We calculate the expected success rate as follows. For each burst, we find the prior probability that the position is contained within the \ac{P48} fields that we observed. We then compute the fraction of afterglows from Kann's sample (which has a mean and standard deviation of $22 \pm 2$~mag at $t = 1$~day) that are brighter than $R = 20.6$\,mag at the same age as when the \ac{P48} observations started. The product of these two numbers is the prior probability of detection for that burst. By summing over all of the \ac{IPTF}/\ac{GBM} bursts, we obtain the expected number of detections. Within 95\% confidence bootstrap error bars, we find an expected 5.5\nobreakdashes--8.5 detections, or a success rate of 16\%\nobreakdashes--24\%. This is consistent with the actual success rate of 23\%.

This suggests that the success rate is currently limited by the survey area and the response time (dictated by sky position and weather). We could increase the success rate by decreasing the maximum time since trigger at which we begin follow\nobreakdashes-up. We could increase the success rate without adversely affecting the number of detections by simply searching a greater area for coarsely localized events.

\ifinthesis
\else
\figsetstart
\figsetnum{3}
\figsettitle{Light curves and SEDs}

\figsetgrpstart
\figsetgrpnum{3.1}
\figsetgrptitle{GRB~130702A~/~iPTF13bxl}
\figsetplot{GRB130702A_sed}
\figsetgrpnote{Light curve and \ac{SED} of GRB~130702A~/~iPTF13bxl.}
\figsetgrpend

\figsetgrpstart
\figsetgrpnum{3.2}
\figsetgrptitle{GRB~131011A~/~iPTF13dsw}
\figsetplot{GRB131011A_sed}
\figsetgrpnote{Light curve of GRB~131011A~/~iPTF13dsw.}
\figsetgrpend

\figsetgrpstart
\figsetgrpnum{3.3}
\figsetgrptitle{GRB~131231A~/~iPTF13ekl}
\figsetplot{GRB131231A_sed}
\figsetgrpnote{Light curve and \ac{SED} of GRB~131231A~/~iPTF13ekl.}
\figsetgrpend

\figsetgrpstart
\figsetgrpnum{3.4}
\figsetgrptitle{GRB~140508A~/~iPTF14aue}
\figsetplot{GRB140508A_sed}
\figsetgrpnote{Light curve and \ac{SED} of GRB~140508A~/~iPTF14aue.}
\figsetgrpend

\figsetgrpstart
\figsetgrpnum{3.5}
\figsetgrptitle{GRB~140606B~/~iPTF14bfu}
\figsetplot{GRB140606B_sed}
\figsetgrpnote{Light curve and \ac{SED} of GRB~140606B~/~iPTF14bfu.}
\figsetgrpend

\figsetgrpstart
\figsetgrpnum{3.6}
\figsetgrptitle{GRB~140620A~/~iPTF14cva}
\figsetplot{GRB140620A_sed}
\figsetgrpnote{Light curve and \ac{SED} of GRB~140620A~/~iPTF14cva.}
\figsetgrpend

\figsetgrpstart
\figsetgrpnum{3.7}
\figsetgrptitle{GRB~140623A~/~iPTF14cyb}
\figsetplot{GRB140623A_sed}
\figsetgrpnote{Light curve of GRB~140623A~/~iPTF14cyb.}
\figsetgrpend

\figsetgrpstart
\figsetgrpnum{3.8}
\figsetgrptitle{GRB~140808A~/~iPTF14eag}
\figsetplot{GRB140808A_sed}
\figsetgrpnote{Light curve and \ac{SED} of GRB~140808A~/~iPTF14eag.}
\figsetgrpend

\figsetend

\begin{figure*}
\centering
\includegraphics[height=\textheight]{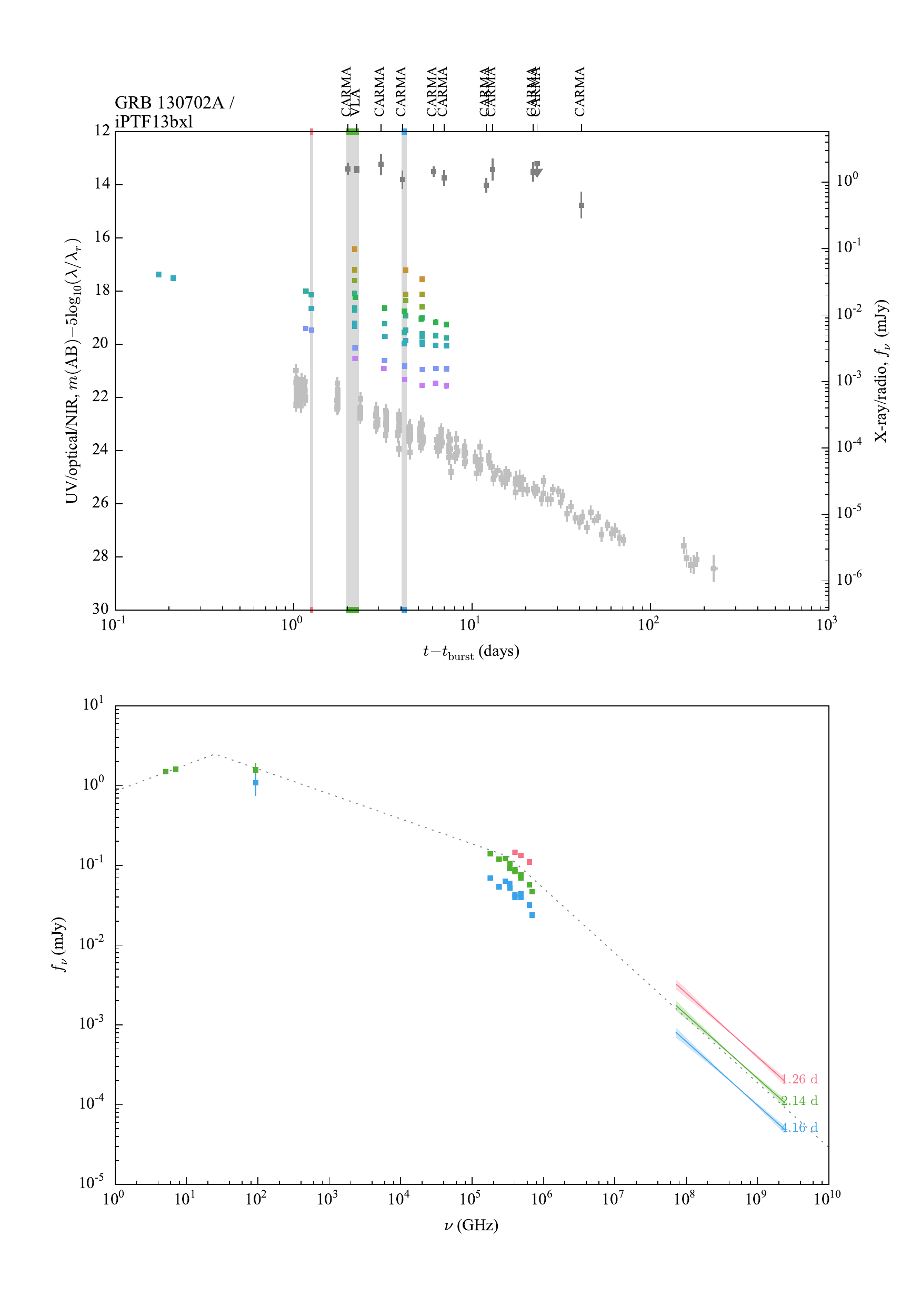}
\caption{\label{fig:lcs-seds}Light curves and \acp{SED} of \ac{GBM}\nobreakdashes--\ac{IPTF} afterglows; GRB~130702A~/~iPTF13bxl is shown here in the print version. \\ 
(An extended version of this figure set is available in the online journal.)}
\end{figure*}
\fi

Over the next few sections, we summarize the observations and general physical interpretation of all of the \ac{GBM}\nobreakdashes--\ac{IPTF} afterglows detected to date. Figure~\ref{fig:lcs-seds} shows the light curves and \acp{SED} spanning X\nobreakdashes-ray, UV, optical, IR, and radio frequencies. Table~\ref{table:spectra} contains a log of our spectroscopic observations.
Table~\ref{tab:photometry} lists a selection of ultraviolet, optical, and infrared observations, including all of our \ac{P48} and \ac{P60} observations. Table~\ref{tab:radio} lists all of our radio detections.

\ifinthesis
\begin{deluxetable*}{llllp{2.5cm}l}
\tabletypesize{\footnotesize}
\else
\begin{deluxetable*}{llllll}
\fi
\tablewidth{\textwidth}
\tablecaption{\label{table:spectra}Log of Spectroscopic Observations}
\tablehead{
    \colhead{Date} &
    \colhead{Telescope} &
    \colhead{Instrument} &
    \colhead{Wavelengths (\AA)} &
        \colhead{Lines} &
    \colhead{References}
}
\startdata
\cutinhead{GRB~131011A/iPTF13dsw}
2013~Oct~12~08:56 &
Gemini South &
GMOS &
5100--9300 &
none &
\citet{GCN15324} \\
2013~Oct~13~03:59 &
ESO/VLT UT3 &
X-shooter &
3100--5560 &
Ly$\alpha$, \ion{Si}{2}, \ion{C}{2}, \ion{C}{4}, \ion{Al}{2} &
\citet{GCN15330} \\
\nodata &
\nodata &
\nodata &
5550--10050 &
\ion{Fe}{2}, \ion{Mg}{2} &
\nodata \\
\cutinhead{GRB~131231A/iPTF13ekl}
2014~Jan~01~02:15 &
Gemini South &
GMOS &
6000--10000 &
[\ion{O}{2}], [\ion{O}{3}], \ion{Ca}{2} H+K &
\citet{GCN15652} \\
\cutinhead{GRB~140508A/iPTF14aue}
2014~May~08~18:55 &
HCT &
HFOSC &
3800--8400 &
\ion{Fe}{2}, \ion{Mg}{2} &
\citet{GCN16244} \\
2014~May~09 06:33 &
APO &
DIS &
3200--9800 &
none &
none \\
\cutinhead{GRB~140606B/iPTF14bfu}
2014~Jun~07~19:16 &
Keck~II &
DEIMOS &
4500--9600 &
[\ion{O}{2}], [\ion{O}{3}], H$\alpha$, \ion{Ca}{2} H+K &
\citet{GCN16365} \\
\cutinhead{GRB~140620A/iPTF14cva}
2014~Jun~20~14:00 &
Gemini North &
GMOS &
5090--9300 &
\ion{Mg}{1}, \ion{Mg}{2}, \ion{Fe}{2}, \ion{Al}{2}, \ion{Si}{2}, \ion{Si}{2}$^*$ &
\citet{GCN16425} \\
\nodata &
\nodata &
\nodata &
4000--6600 &
\nodata &
\nodata \\
\cutinhead{GRB~140623A/iPTF14cyb}
2014~Jun~23~08:10 &
Gemini North &
GMOS &
4000--6600 &
\ion{Mg}{2}, \ion{Fe}{2}, \ion{Al}{2}, \ion{Si}{2}, \ion{Al}{3}, \ion{C}{1}, \ion{C}{4} &
\citet{GCN16442} \\
\cutinhead{GRB~140808A/iPTF14eag}
2014~Aug~08~21:43 &
GTC &
OSIRIS &
3630--7500 &
DLA, \ion{S}{2}, \ion{Si}{2}, \ion{O}{1}, \ion{C}{2}, \ion{Si}{4}, \ion{Fe}{2}, \ion{Al}{2}, \ion{C}{4} &
\citet{GCN16671}
\enddata
\end{deluxetable*}

\begin{figure*}
    \centering
    \includegraphics[width=\textwidth]{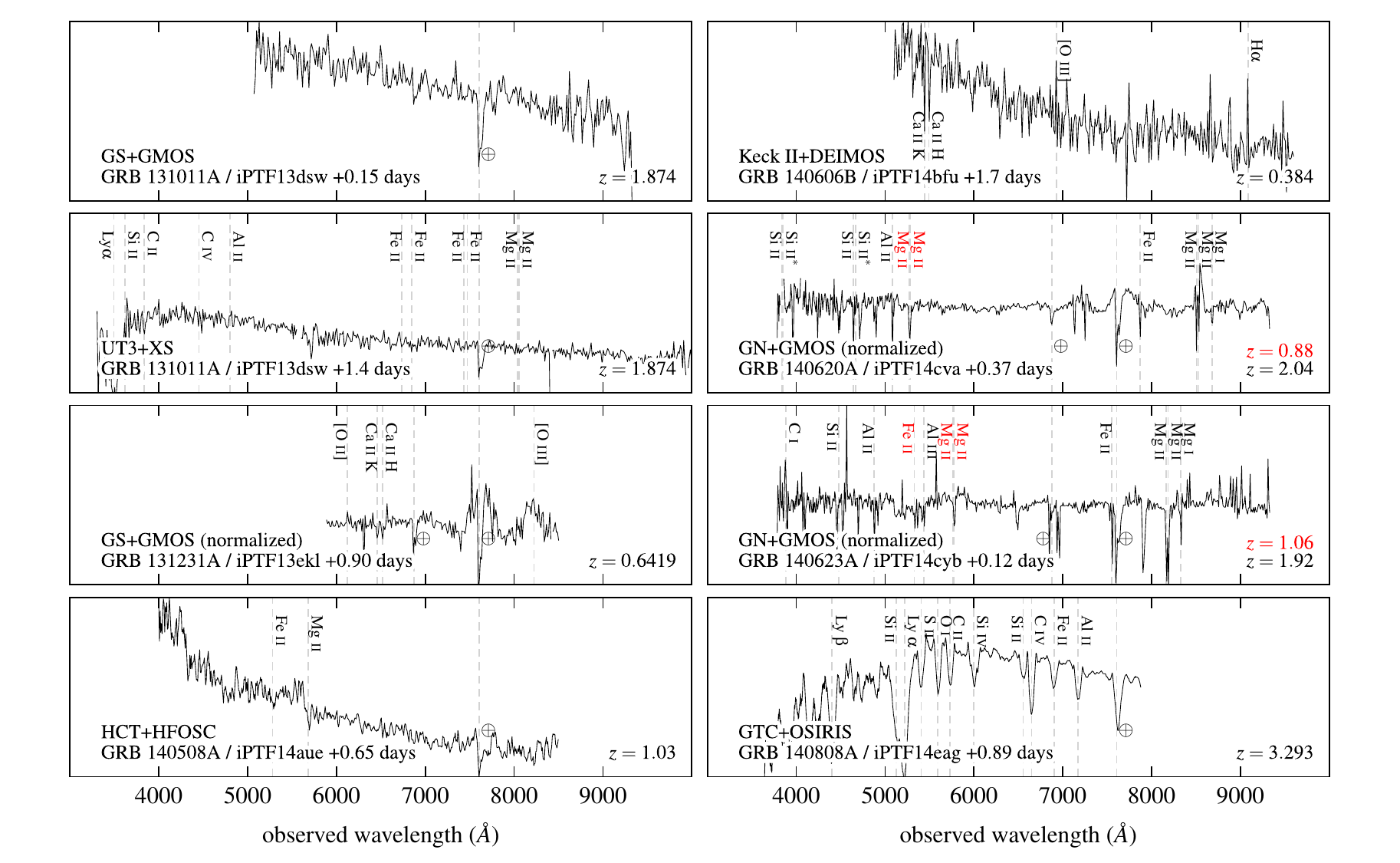}
    \caption[Afterglow spectra]{\label{fig:spectra}Afterglow spectra. The horizontal axis shows wavelength in vacuum in the observer frame, and the vertical axis shows scaled flux. Lines at the redshift of the putative host are labeled in black; lines corresponding to any intervening absorbing systems are labeled in red. Note that in cases where one or fewer lines are discernible in our spectra, the redshifts have been reported in \acp{GCN} by other groups.}
\end{figure*}

\ifinthesis
\else
\makeatletter{}\begin{deluxetable}{lrrr@{\enspace}r@{\enspace}l@{$\,\pm\,$}l}
\tablewidth{\columnwidth}
\tablecolumns{7}
\tablecaption{\label{tab:photometry}Optical Observations of \acs{GBM}--\acs{IPTF} Afterglows}
\tablehead{
    \colhead{Date (mid)} & \colhead{Inst.\tablenotemark{a}} & \colhead{$\Delta t$\tablenotemark{b}} & \multicolumn{4}{c}{Mag.\tablenotemark{c}}
}
\startdata
        \cutinhead{GRB\,130702A/iPTF13bxl} \\
        2013 Jul 02 04:18 &
        P48 &
        0.18 &
        $R$ &
        $=$ &17.38 &
        0.04 \\
        
        2013 Jul 02 05:10 &
        P48 &
        0.21 &
        $R$ &
        $=$ &17.52 &
        0.04 \\
        
        2013 Jul 03 04:13 &
        P60 &
        1.17 &
        $g$ &
        $=$ &18.80 &
        0.04 \\
        
        2013 Jul 03 04:15 &
        P60 &
        1.17 &
        $i$ &
        $=$ &18.42 &
        0.04 \\
        
        2013 Jul 03 06:16 &
        P60 &
        1.26 &
        $i$ &
        $=$ &18.56 &
        0.06 \\
        
        2013 Jul 03 06:17 &
        P60 &
        1.26 &
        $r$ &
        $=$ &18.66 &
        0.05 \\
        
        2013 Jul 03 06:20 &
        P60 &
        1.26 &
        $g$ &
        $=$ &18.86 &
        0.04
\enddata
\tablenotetext{a}{RATIR data are from \citet{GCN14980,GCN14993,GCN16236}. GROND data are from \citet{GCN15328}. Keck near\nobreakdashes-infrared data for \ac{GRB}\,140606B are from \citet{GCN16387}.}
\tablenotetext{b}{Time in days relative to \ac{GBM} trigger.}
\tablenotetext{c}{Magnitudes are in the AB system \citep{ABMags}.}
\tablecomments{A machine readable version of this table is available in the online journal.}
\end{deluxetable}
 
\makeatletter{}\begin{deluxetable}{llrr@{\enspace}r@{\enspace}r@{}c@{}l}
\tablewidth{\columnwidth}
\tablecolumns{8}
\tablecaption{\label{tab:radio}Radio Observations of \acs{GBM}--\acs{IPTF} Afterglows}
\tablehead{
    \colhead{Date (Start)} & Inst.\tablenotemark{a} & \colhead{$\Delta t$\tablenotemark{b}} & \multicolumn{5}{c}{Flux Density\tablenotemark{c}}
}
\startdata
        \cutinhead{GRB\,130702A/iPTF13bxl} \\
        2013 Jul 04 &
        CARMA &
        2 &
        $f_\nu(93)$ &
        
        $=$ &
        1580 &
        $\pm$ &
        330
         \\
        
        2013 Jul 04 &
        VLA &
        2.3 &
        $f_\nu(5.1)$ &
        
        $=$ &
        1490 &
        $\pm$ &
        75
         \\
        
        2013 Jul 04 &
        VLA &
        2.3 &
        $f_\nu(7.1)$ &
        
        $=$ &
        1600 &
        $\pm$ &
        81
         \\
        
        2013 Jul 05 &
        CARMA &
        3.1 &
        $f_\nu(93)$ &
        
        $=$ &
        1850 &
        $\pm$ &
        690
         \\
        
        2013 Jul 06 &
        CARMA &
        4.1 &
        $f_\nu(93)$ &
        
        $=$ &
        1090 &
        $\pm$ &
        350
         \\
        
        2013 Jul 08 &
        CARMA &
        6.1 &
        $f_\nu(93)$ &
        
        $=$ &
        1440 &
        $\pm$ &
        260
         \\
        
        2013 Jul 08 &
        CARMA &
        7 &
        $f_\nu(93)$ &
        
        $=$ &
        1160 &
        $\pm$ &
        320
         \\
        
        2013 Jul 14 &
        CARMA &
        12 &
        $f_\nu(93)$ &
        
        $=$ &
        900 &
        $\pm$ &
        230
         \\
        
        2013 Jul 15 &
        CARMA &
        13 &
        $f_\nu(93)$ &
        
        $=$ &
        1550 &
        $\pm$ &
        590
         \\
        
        2013 Jul 24 &
        CARMA &
        22 &
        $f_\nu(93)$ &
        
        $=$ &
        1430 &
        $\pm$ &
        480
         \\
        
        2013 Jul 25 &
        CARMA &
        23 &
        $f_\nu(93)$ &
        
        $<$ &
        1890 &
        &
         \\
        
        2013 Aug 12 &
        CARMA &
        41 &
        $f_\nu(93)$ &
        
        $=$ &
        450 &
        $\pm$ &
        210
        
\enddata
\tablenotetext{a}{The \ac{ATCA} observation is from \citet{GCN15395}.}
\tablenotetext{b}{Time in days relative to \ac{GBM} trigger.}
\tablenotetext{c}{Flux density in $\mu$Jy as a function of frequency in GHz. For detections, the confidence intervals are $1\sigma$ statistical uncertainties added in quadrature with an estimated 5\% systematic error. For non\nobreakdashes-detections, we show $3\sigma$ upper limits.}
\tablecomments{A machine readable version of this table is available in the online journal.}
\end{deluxetable}
 
\fi

\subsection{\ac{GRB}\,130702A/iPTF13bxl}

This is the first \ac{GBM} burst whose afterglow we discovered with iPTF \citep{iPTF13bxl}, indeed the first afterglow ever to be pinpointed based solely on a \emph{Fermi} \ac{GBM} localization. It is also the lowest\nobreakdashes-redshift \ac{GRB} in our sample, so it has the richest and most densely sampled broadband afterglow data. It has two other major distinctions: its associated \ac{SN} (\ac{SN}~2013dx, \citealt{GCN14994,GCN14996,GCN14998,GCN15000}) was detected spectroscopically, and its prompt energetics are intermediate between \acp{llGRB} and standard cosmic bursts (see below).

Based on the \emph{Fermi} \ac{GBM} ground localization with an error radius of $4\arcdeg$, we imaged 10 fields twice with the \ac{P48} at $\Delta t = t - t_\mathrm{GBM} = 4.2$~hr after the burst.\footnote{At the time, our tiling algorithm selected fields based on an empirical calibration of \emph{Fermi} \ac{GBM}'s systematic errors. We had selected bursts that were detected by both \emph{Swift} and \emph{Fermi} and constructed a fit to a cumulative histogram of the number of bursts whose \ac{BAT} or \ac{XRT} positions were within a given number of nominal $1\sigma$ statistical radii of the center of the \emph{Fermi} error circle. Our tiling algorithm scaled this fit by the $1\sigma$ radius of the burst in question and then constructed a 2D angular probability distribution from it. For sufficiently large error radii, this prescription produced probability distributions that had a hole in the middle. For this reason, the tiling algorithm picked out \ac{P48} fields that formed an annulus around the \ac{GBM} 1$\sigma$ error circle (not, as we stated in \citealt{iPTF13bxl}, because of a lack of reference images).} We scheduled P60 imaging and P200 spectroscopy for three significantly varying sources. Of the three, iPTF13bxl showed the clearest evidence of fading in the \ac{P48} images. Its spectrum at $\Delta t = 1.2$~days consisted of a featureless blue continuum. We triggered \emph{Swift}, which found a bright X-ray source at the position of iPTF13bxl \citep{GCN14967,GCN14973}. Shortly after we issued our \ac{GCN} circular \citep{GCN14967}, \citet{GCN14971} announced that the burst had entered the FOV of LAT at $\Delta t = 250$~s. The LAT error circle had a radius of $0\fdg5$, and its center was $0\fdg8$ from iPTF13bxl. An \ac{IPN} triangulation with \emph{MESSENGER} (GRNS), \emph{INTEGRAL} (SPI\nobreakdashes-ACS), \emph{Fermi}\nobreakdashes-\ac{GBM}, and Konus\nobreakdashes-\emph{Wind} \citep{GCN14974} yielded a $0\fdg46$\nobreakdashes-wide annulus that was also consistent with the OT.

The afterglow's position is $0\farcs 6$ from an $R = 23.01$\,mag source that is just barely discernible in the \ac{P48} reference images. A spectrum from NOT+ALFOSC \citep{GCN14983} determined a redshift of $z = 0.145$ for a galaxy $7\farcs 6$ to the south of iPTF13bxl. At $\Delta t = 2.0$~days, we obtained a Magellan+IMACS spectrum \citep{GCN14985} and found weak emission lines at the location of the afterglow that we interpreted as H$\alpha$ and [\ion{O}{3}] at the same redshift. \citet{13bxlhost} characterized the burst's host environment in detail and concluded that it exploded in a dwarf satellite galaxy.

Joining the two \ac{P48} observations at $\Delta t < 1$~day to the late\nobreakdashes-time P60 light curve requires a break at $\Delta t = 1.17 \pm 0.09$~days, with slopes $\alpha_{\mathrm{O},1}=0.57 \pm 0.03$ and $\alpha_{\mathrm{O},2}=1.05 \pm 0.03$ before and after the break, respectively. The XRT light\nobreakdashes-curve begins just prior to this apparent break and seems to follow the late\nobreakdashes-time optical decay (until the \ac{SN} begins to dominate at $\Delta t = 5$~days), although the automated \emph{Swift} light curve analysis \citep{2009MNRAS.397.1177E} also suggests a possible X\nobreakdashes-ray break with about the same time and slopes. This hints at an achromatic break, normally a signature of a jet. However, the late slope and the change in slope are both unusually shallow for a jet break. Furthermore, the radio light curve does not exhibit a break. The change in slope is also a little too large for cooling frequency crossing the band (for which one would expect $\Delta \alpha = 1/4$). An energy injection or a structured jet model may provide a better fit \citep{ModelsForAchromaticLightCurveBreaks}.

Late\nobreakdashes-time $\Delta t > 1$~day observations include several P60 $gri$ observations, three RATIR $r'i'ZYJH$ epochs, an extensive \emph{Swift} XRT and UVOT light curve, and radio observations with VLA and \ac{CARMA} (although of the VLA data, we only have access to the first observation). The optical and X-ray spectral slopes are similar, $\beta_\mathrm{O} = 0.7 \pm 0.1$ and $\beta_\mathrm{X} = 0.8 \pm 0.1$. An SED at $2 < \Delta t < 2.3$~days is well explained by the standard external shock model \citep{AfterglowSpectra} in the slow cooling regime, with $\nu_m$ lying between the VLA and \ac{CARMA} frequencies and $\nu_c$ in the optical. This fit requires a relatively flat electron spectrum, $d n_\mathrm{e} / d \gamma_\mathrm{e} \propto {\gamma_\mathrm{e}}^{-p}$ with $p \approx 1.6$, cut off at high energies. Applying the relevant closure relations (for the case of $1 < p < 2$, see \citealt{AfterglowSpectraFlatSpectrum}) to $\alpha_X$ and $\beta_X$ permits either an ISM or wind environment.

Our late\nobreakdashes-time spectroscopy and analysis of the \ac{SN} will be published separately (S.~B.~Cenko~et~al. 2015, in~preparation).

\subsection{\ac{GRB}\,131011A/iPTF13dsw}

We started \ac{P48} observations of \emph{Fermi} trigger 403206457 \citep{GCN15331} about 11.6\,hr after the burst. The optical transient iPTF13dsw \citep{GCN15324} faded from $R=19.7$\,mag to $R=20.2$\,mag from 11.6~to~14.3\,hr. The latest pre\nobreakdashes-trigger image on 2013~September~25 had no source at this location to a limit of $R > 20.6$~mag. The optical transient continued to fade as it was monitored by several facilities \citep{GCN15325,GCN15326,GCN15327,GCN15328,GCN15341}.

At 15.1\,hr after the burst, we obtained a spectrum of iPTF13dsw with the \acf{GMOS} on the Gemini\nobreakdashes--South telescope. \ac{GMOS} was configured with the R400 grating with a central wavelength of 7200\,{\AA} and the 1\arcsec slit, providing coverage over the wavelength range of 5100\nobreakdashes--9300\,{\AA} with a resolution of $\approx 3$\,{\AA}.  No prominent features were detected over this bandpass, while the spectrum had a typical SNR of $\approx 3$\nobreakdashes--4 per 1.4\,{\AA} pixel. \citet{GCN15330} observed the optical transient with the X\nobreakdashes-Shooter instrument on the ESO 8.2\nobreakdashes-m \ac{VLT}. In their spectrum extending from $\sim$3000~to~$\sim24000$\AA, they identified several weak absorption lines from which they derived a redshift of $z = 1.874$. Both spectra are shown in Figure~\ref{fig:spectra}.

The source was detected by \emph{Swift} XRT \citep{GCN15329}, but with insufficient photons for spectral analysis. The source was observed with \ac{ATCA}, but no radio emission was detected. Largely because in our sample this is the oldest afterglow at the time of discovery, there are not enough broadband data to constrain the blast wave physics.

\subsection{\ac{GRB}\,131231A/iPTF13ekl}

\ac{GRB}\,131231A was detected by \emph{Fermi} \ac{LAT} \citep{GCN15640} and \ac{GBM} \citep{GCN15644}, with photons of energies up to 9.7~GeV. \citet{GCN15641} observed the \ac{LAT} error circle with the 1\nobreakdashes-m telescope at Mt. Nanshan, Xinjiang, China. At 7.9\,hr after the burst, they detected a single $R = \sim17.6$~mag source that was not present in SDSS images. At 17.3\,hr after the burst, \citet{GCN15642} observed the afterglow candidate with the MOSaic CAmera (MOSCA) on the 2.56\nobreakdashes-m Nordic Optical Telescope (NOT). The source had faded to $R = 18.6$.

Although we had imaged 10 \ac{P48} fields shortly after the \emph{Fermi} trigger \citep{GCN15643}, due to the short visibility window at Palomar we were only able to obtain one epoch. At 1.45\,hr after the burst, we detected an $R = 15.7$\,mag optical transient iPTF13ekl at the position of the Nanshan candidate. Though our single detection of iPTF13ekl could not by itself rule out that the source was a moving solar system object, the Nanshan detection at 6.46\,hr, fitting a decay with a power\nobreakdashes-law index of $\alpha = 1.03$, was strong evidence that the transient was the optical afterglow of \ac{GRB}\,131231A.

On January 1.09 UT (21.5\,hr after the trigger), we observed the afterglow with Gemini South using the \ac{GMOS} camera \citep{GMOS} in Nod\&Shuffle mode: we obtained 32 dithered observations of 30\,s each at an average airmass of 2. We analyzed this data set using the dedicated GEMINI package under the IRAF environment and extracted the 1-dimensional spectrum using the APALL task. We determined the redshift of the \ac{GRB}, based on the simultaneous identification of forbidden nebular emission lines ([\ion{O}{2}], [\ion{O}{3}]) and absorption features (CaH\&K) at the same redshift of $z=0.6419$. In Figure~\ref{fig:spectra}, we show the normalized spectrum.

The source was also detected by \emph{Swift} XRT \citep{GCN15648} and UVOT \citep{GCN15673}, as well as \ac{CARMA} \citep{GCN15680}.

With only the millimeter, optical, and X\nobreakdashes-ray observations, the SED is highly degenerate. Contributing to the degeneracy, the X\nobreakdashes-ray and optical observations appear to fall on the same power\nobreakdashes-law segment. It is consistent with either fast or slow cooling if the greater of $\nu_c$ or $\nu_m$ is near the optical, assuming a flat electron distribution with $p \sim 1.5$. It is also consistent with slow cooling if $\nu_c$ is above the X\nobreakdashes-ray band and $p \sim 2.6$.

\subsection{\ac{GRB}\,140508A/iPTF14aue}

This burst was detected by \emph{Fermi} \ac{GBM} and \emph{INTEGRAL} SPI\nobreakdashes-ACS \citep{GCN16224}, as well as by Konus\nobreakdashes-\emph{Wind}, Mars Odyssey (not included in the \ac{GCN} circular), \emph{Swift} \ac{BAT} (outside the coded \ac{FOV}), and \emph{MESSENGER}, yielding a $1 \fdg 5 \times 12\arcmin$ \ac{IPN} error box \citep{GCN16225}.

Due to poor weather early in the night, \ac{P48} observations started 6.7\,hr after the trigger \citep{GCN16226}. We found one optical transient candidate within the \ac{IPN} triangulation, iPTF14aue, which faded from $r = 17.89 \pm 0.01$~mag with a power\nobreakdashes-law fit of $\alpha = 1.12 \pm 0.1$ over a timescale of 1.5\,hr.

We triggered a \emph{Swift} \ac{TOO}. From 0.8~to~8.1~days after the trigger, \emph{Swift} XRT detected a coincident X\nobreakdashes-ray source that faded with a power law $\alpha = 1.48\;(+0.15, -0.14)$ \citep{GCN16232,GCN16254}. The source was also detected by \emph{Swift} UVOT \citep{GCN16243}.

\citet{GCN16228} obtained a 20\nobreakdashes-minute, 3800\nobreakdashes--7200\,{\AA} spectrum of iPTF14aue with the 6\nobreakdashes-m BTA telescope in Zelenchukskaia. Exhibiting no absorption features, this established an upper limit of $z < 2.1$. \citet{GCN16229} used the Andalucia Faint Object Spectrograph and Camera (ALFOSC) on NOT to get an 1800\,s spectrum spanning 3200\nobreakdashes--9100\,{\AA}, and found several absorption features at redshift $z = 1.03$. Consistent redshifts were reported by \citet{GCN16231} with the ACAM instrument on the 4.2\nobreakdashes-m William Herschel Telescope and by \citet{GCN16244} with \ac{HFOSC} on the 2\nobreakdashes-m \ac{HCT}. This last spectrum is shown in Figure~\ref{fig:spectra}.

Due to the brightness of the optical transient, optical photometry was available from several facilities up to 4.5 days after the burst \citep{GCN16227,GCN16228,GCN16229,GCN16235,GCN16236,GCN16246,GCN16259,GCN16260}.

\citet{GCN16266} detected the source with \ac{VLA} 5.2 days after the Fermi trigger, at 6.1 GHz ($C$ band) and at 22 GHz ($K$ band). A broadband \ac{SED} constructed from P60 and \ac{XRT} data from around this time is consistent with $p \approx 2$. Because $p$ is not distinguishable from 2, we cannot discriminate between fast and slow cooling based on this one time slice. However, given the late time of this observation, the slow cooling interpretation is more likely, putting $\nu_m$ between the radio and optical bands and $\nu_c$ between the optical and X\nobreakdashes-ray. Because the \ac{VLA} light curve is decreasing with time, an \ac{ISM} circumburst density profile is favored.

\subsection{\ac{GRB}\,140606B/iPTF14bfu}

\emph{Fermi} trigger 423717114 \citep{GCN16363} was observable from Palomar for several hours, starting about 4.3\,hr after the time of the burst. Based on the final \ac{GBM} localization, we searched ten \ac{P48} fields and found several plausible optical transient candidates \citep{GCN16360}.

iPTF14bfu had no previous detections in iPTF between 2013~May~23 and October~13. Its position was outside the SDSS survey footprint, but it had no plausible host associations in VizieR \citep{VizieR}. From 4.3 to 5.5\,hr after the burst, it faded from $R = 19.89 \pm 0.10$ to $20.32 \pm 0.14$ mag, fitting a power law of $\alpha = -1.6 \pm 0.7$ relative to the time of the \ac{GBM} trigger. iPTF14bfw ($R = 19.96 \pm 0.06$ mag) was coincident with an $r = 21.27$ galaxy in SDSS DR10 and displayed no statistically significant photometric variation over the course of our \ac{P48} observations. iPTF14bgc ($R = 18.44 \pm 0.02$ mag) was coincident with an $R = 21.07 \pm 0.08$ mag point source in our co\nobreakdashes-added reference image composed of exposures from 2013~July~31 through September~24. iPTF14bga ($R = 19.75 \pm 0.06$ mag) was likewise coincident with a $R = 20.42 \pm 0.17$ mag point source in our reference image composed of exposures from 2011~July~29 through October~20.

On the following night, we observed all four candidates again with \ac{P48} and P60 \citep{GCN16362}. iPTF14bfw and iPTF14bga had not faded relative to the previous night. iPTF14bgc had faded to $R = 20.68 \pm 0.21$ mag, consistent with the counterpart in our reference images but significantly fainter than the previous night. A power\nobreakdashes-law fit to the decay gave a temporal index of $\alpha = -1.1 \pm 0.1$, entirely consistent with typical \ac{GRB} afterglows. iPTF14bfu was not detected in our \ac{P48} images to a limiting magnitude of $R < 21.1$, but it was detected in stacked P60 images ($r = 21.1 \pm 0.2$), consistent with a power law of $\alpha \sim -0.5$.

An \ac{IPN} triangulation from \emph{Fermi}, Konus\nobreakdashes--\emph{Wind}, and MESSENGER yielded a long, slender $14\fdg18 \times 0\fdg414$ error box that contained iPTF14bfu and iPTF14bfw \citep{GCN16369}.

We obtained two 900~s spectra with the DEIMOS spectrograph on the Keck~II 10\,m telescope \citep{GCN16365}. On a blue continuum, we found [\ion{O}{2}], [\ion{O}{3}], and H\,$\alpha$ emission features and \ion{Ca}{2} absorption features, at a common redshift of $z = 0.384$. A galaxy offset by $\sim 2\arcsec$ along the slit showed the same emission lines at the same redshift.

\emph{Swift} XRT observed the location of iPTF14bfu for a total of 9~ks from 2.1~to~9.3 days after the \ac{GBM} trigger, and found a source that faded with a power\nobreakdashes-law fit of $\alpha = -1.0\;(+0.7,-0.6)$ \citep{GCN16366,GCN16373,GCN16412}.

At 18.4 days after the trigger, we obtained a 1200~s spectrum of iPTF14bfu with the Low Resolution Imaging Spectrometer (LRIS) on the Keck~I 10\nobreakdashes-meter telescope \citep{GCN16454}. The spectrum had developed broad emission features. A comparison using Superfit \citep{Superfit} showed a good match to \ac{SN}~1998bw near maximum light, indicating that the source had evolved into an \ac{SNIcBL}. Our late\nobreakdashes-time photometry and spectroscopy will published separately \citep{iPTF14bfuSN}.

Although there were three radio detections of \ac{GRB}\,140606B, only during the first \ac{CARMA} detection does the optical emission appear to be dominated by the afterglow. We can construct an \ac{SED} around this time using nearly coeval \ac{DCT} and \ac{XRT} data. Because of the faintness of the X\nobreakdashes-ray afterglow, the spectral slopes $\beta_\mathrm{X}$ and $\beta_\mathrm{OX}$ are only weakly determined. As a result, there is a degeneracy between two plausible fits. The first has $\nu_m$ anywhere below the \ac{CARMA} band, $\nu_c$ just below the X\nobreakdashes-rays, and $p \approx 2$. The second has $\nu_m$ just above the radio and $\nu_c$ in the middle of the \ac{XRT} band, with $p \approx 2.2$.

The early \ac{P48} observations do not connect smoothly with the P60 and \ac{DCT} observations from $\Delta t = 1$~to~4~days. This may indicate a steep\nobreakdashes--shallow\nobreakdashes--steep time evolution requiring late\nobreakdashes-time energy injection, or may just indicate that the afterglow is contaminated by light from the host galaxy or the \ac{SN} at relatively early times.

\subsection{\ac{GRB}\,140620A/iPTF14cva}

This burst is distinctive in our sample for two reasons. First, it is the earliest afterglow detection in the \ac{IPTF} sample at $\Delta t = 0.25$\,hr. Second, its broadband \ac{SED} is not clearly explainable by the standard forward shock model.

\emph{Fermi} trigger 424934131 \citep{GCN16426} was observable from Palomar for about 6\,hr from the time of the burst. Based on the ground localization, we started observing ten \ac{P48} fields about 10 minutes after the trigger. Based on the final localization, we added 10 more fields, for a total of 20, about an hour after the trigger.

The candidate iPTF14cva \citep{GCN16425} was contained within one of the early 10 fields. From 14.9 to 87.2 minutes after the trigger, the candidate faded from $R = 17.60 \pm 0.01$ to $18.80 \pm 0.02$~mag, consistent with a somewhat slow power law of $\alpha = 0.62 \pm 0.01$.

We observed the candidate with GMOS on the 8\nobreakdashes-m Gemini North telescope. Starting 8.8\,hr after the trigger, we obtained two 900~s spectra extending from 4000 to 9300\,{\AA}. We detected \ion{Mg}{2} and \ion{Fe}{2} absorption lines at $z = 0.88$ and many absorption features at a common redshift of $z = 2.04$. The lack of Ly$\alpha$ absorption implied an upper limit of $z \sim 2.3$ and suggested that $z = 2.04$ was the redshift of the source.

We triggered \emph{Swift} and \ac{VLA} follow\nobreakdashes-up. In a 3~ks exposure starting 10.4\,hr after the \emph{Fermi} trigger, \emph{Swift} XRT detected an X\nobreakdashes-ray source with a count rate of $1.2 \times 10^{-1}$~counts\,s$^{-1}$ \citep{GCN16428}. Over the next four days of \emph{Swift} observations, the X-ray source faded with a slope $\alpha = 1.32 \pm 0.16$ \citep{GCN16455}. A fading source was also detected by \emph{Swift} UVOT \citep{GCN16432}.

The source was detected by \ac{VLA} on June 23 at 6.1~GHz ($C$ band) at $108 \pm 15$~$\mu$Jy and at 22~GHz ($K$ band) at $62 \pm 15$~$\mu$Jy.  On June 30, there was a marginal detection in $C$ band with $48 \pm 12$~$\mu$Jy and no detection in $K$ band with a noise level of 15~$\mu$Jy rms. 

The optical transient was also observed in $R$ band by the Konkoly Observatory \citep{GCN16440} and the 1\nobreakdashes-m telescope at the Tien Shan Astronomical Observatory \citep{GCN16453}.

The \ac{SED} of this afterglow cannot be explained by a standard forward shock model. If we place the peak frequency near the radio band, the optical and X\nobreakdashes-ray fluxes are drastically underpredicted, whereas if we place the peak frequency between the optical and X\nobreakdashes-ray bands, we miss the radio observations by orders of magnitude. This seems to require an additional component. One possibility is that there is a forward shock peak in the UV and a reverse shock peak at low frequencies (similar to \ac{GRB}\,130427A; see \citealt{LaskarGRB130427A,PerleyGRB130427A}). Another possibility is that there is an inverse Compton peak in the UV (similar to \ac{GRB}\,120326A; \citealt{UrataGRB120326A}).

\subsection{\ac{GRB}\,140623A/iPTF14cyb}

\emph{Fermi} trigger 425193729 \citep{GCN16450} was observable from Palomar for about 6\,hr from the time of the burst. Based on the ground localization, we started imaging 10 fields 11 minutes after the trigger. The final \emph{Fermi} localization, which was avilable 2.6\,hr later, shifted by 13\fdg4. Due to the large change in the localization, we calculated only a 4\% chance that the source was contained within the \ac{P48} fields.

Candidate iPTF14cyb \citep{GCN16425}, situated at an extreme edge of the \ac{P48} tiling, was within the 1$\sigma$ confidence region for both the ground and final localizations. From 16 to 83 minutes after the trigger, the source faded from $R = 18.04 \pm 0.01$ to $19.69 \pm 0.06$~mag, consistent with a power-law decay with an index $\alpha = 0.94 \pm 0.03$.

Starting 2.8\,hr after the trigger, we obtained two 900~s GMOS spectra extending from 4000 to 9300\,{\AA}. We detected \ion{Mg}{2} and \ion{Fe}{2} absorption lines at $z = 1.06$ and many absorption features at $z = 1.92$. The lack of Ly$\alpha$ absorption implied that this was the redshift of the burst.

We triggered \emph{Swift}, \ac{VLA}, and \ac{CARMA} follow\nobreakdashes-up. In a 3~ks exposure starting 10.7\,hr after the burst, \emph{Swift} XRT detected an uncataloged X\nobreakdashes-ray source with a count rate of $(2.2 \pm 0.6) \times 10^{-3}$~counts\,s$^{-1}$ \citep{GCN16451}. By 79\,hr after the trigger, the source was no longer detected in a 5~ks exposure \citep{GCN16464}. No radio source was detected with \ac{VLA} in $C$ band (6.1~GHz) to an rms level of 17~$\mu$Jy, or in $K$ band (22~GHz) to an rms level of 18~$\mu$Jy.

Because of the lack of radio detections and the extreme faintness of the X\nobreakdashes-ray afterglow, the broadband behavior of the afterglow does not constrain the shock physics.

\subsection{\ac{GRB}\,140808A/iPTF14eag}

\emph{Fermi} trigger 429152043 \citep{GCN16669} was observable from Palomar about 3\,hr after the burst. We imaged 13 fields with \ac{P48} and found one promising optical transient. iPTF14eag was situated on the extreme edge of one of the \ac{P48} tiles that was just outside the \ac{GBM} 1$\sigma$ contour. It faded from $R = 18.91 \pm 0.06$ to $19.29 \pm 0.10$~mag from 3.35~to~4.91\,hr after the trigger and had no archival counterparts in SDSS or in our own reference images.

Due to Hurricane Iselle, we were unable to use our \ac{TOO} programs on Keck or Gemini North. We requested photometric confirmation of the fading from \ac{HCT} \citep{GCN16684}, submitted a \emph{Swift} \ac{TOO}, and sent our \ac{GCN} circular \citep{GCN16668} to encourage others to obtain a spectrum.

\emph{Swift} observed the position of iPTF14eag from 11.6~to~14.4\,hr after the burst \citep{GCN16670}. An X-ray source was detected with a count rate of $1.5\times10^{-2}$~counts\,s$^{-1}$. In a second observation starting 62.2\,hr after the trigger \citep{GCN16682}, the source had faded to below $2.46 \times 10^{-3}$~counts\,s$^{-1}$. No source was detected by UVOT \citep{GCN16672}.

We obtained a spectrum with the OSIRIS instrument \citep{OSIRIS} on \ac{GTC} in $4\times900$~s exposures with a mean epoch of 21.340\,hr after the burst. We used the R1000B grism and a 1" slit, with a resolution of R~1000. We determined a redshift of 3.293 (improved from the one given by \citealt{GCN16671}) through the identification of strong absorption features. The flux\nobreakdashes-calibrated spectrum is shown in Figure~\ref{fig:spectra}.

The source was detected in radio with \ac{VLA} \citep{GCN16694} and \ac{AMI} \citep{GCN16725}. The broadband \ac{SED} around the time of the VLA detection broadly fits a forward shock model but is poorly constrained due to the lack of a contemporaneous X\nobreakdashes-ray detection. The spectral slope between the two VLA bands is somewhat steeper than the standard low\nobreakdashes-frequency value of $\beta = -1/3$, possibly indicating that the radio emission is self\nobreakdashes-absorbed. We obtained 14 \ac{AMI} observations every 2 or 3 days from 2014~August~8 until 2014~September~12. Observations were 2\nobreakdashes--4\,hr in duration. \ac{AMI} first detected the afterglow 4.6 days post-burst. The \ac{AMI} light curve peaked $\sim$10.6~days post\nobreakdashes-burst at 15.7 GHz, which is characteristic of forward shock emission at radio wavelengths \citep{RadioSelectedGRBAfterglows}.

A peculiar feature of the optical light curve is that the P60 $r$- and $i$\nobreakdashes-band observations at $\Delta t \approx 2$~days appears to be inverted, with a rising rather than falling spectral shape, compared to the earlier P60 photometry at $\Delta t \approx 1$~day. However, this feature is within the error bars and may be merely a statistical fluctuation.

This is the highest\nobreakdashes-redshift burst in our sample and also had the weakest prompt emission in terms of the fluence observed by \ac{GBM}.

\ifinthesis
\else
\begin{figure*}
\includegraphics{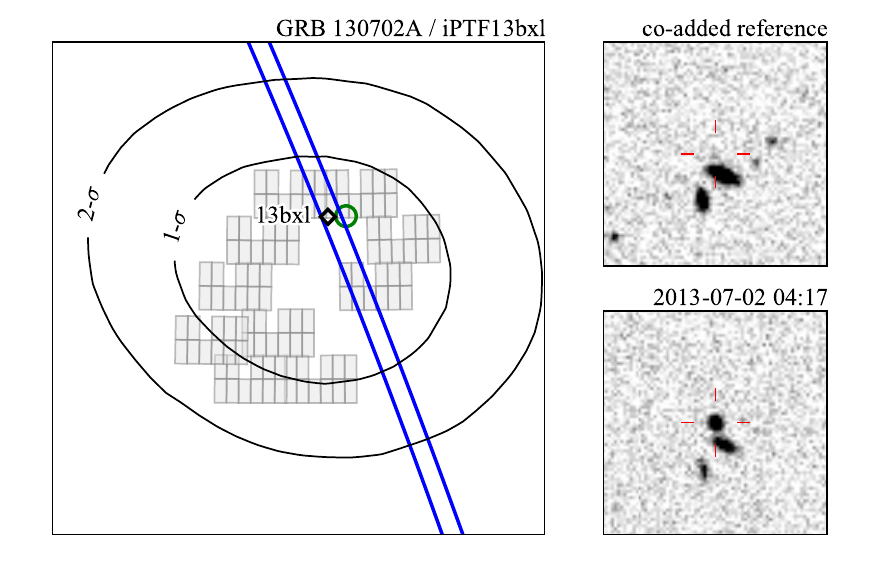}
\includegraphics{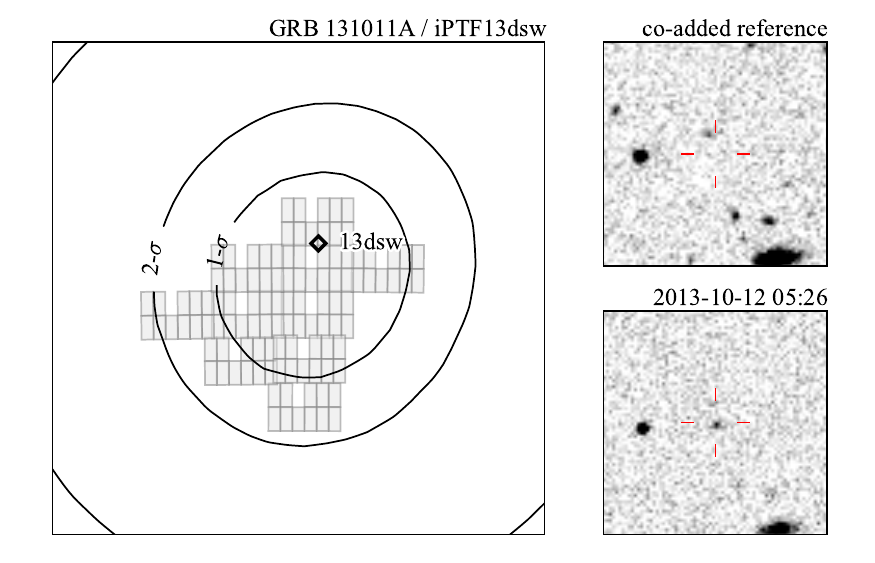}\\
\includegraphics{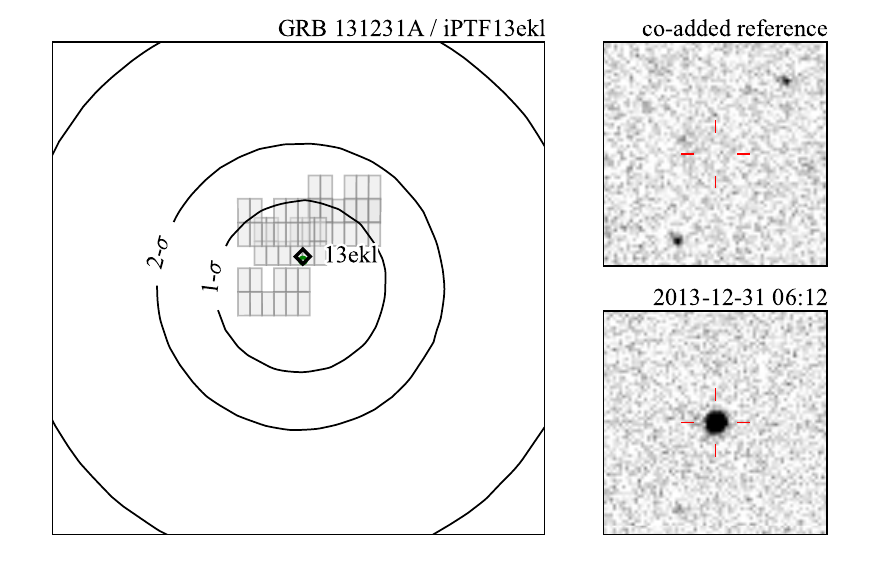}
\includegraphics{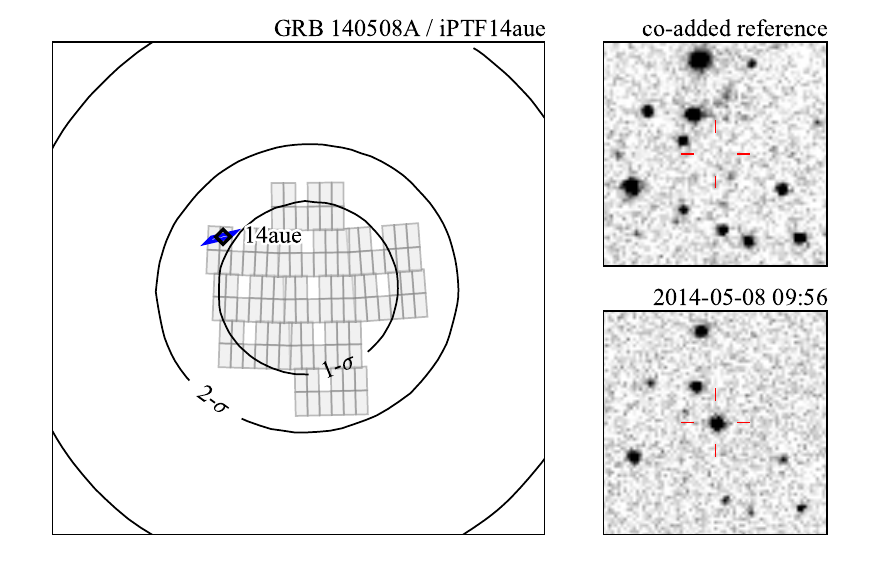}\\
\includegraphics{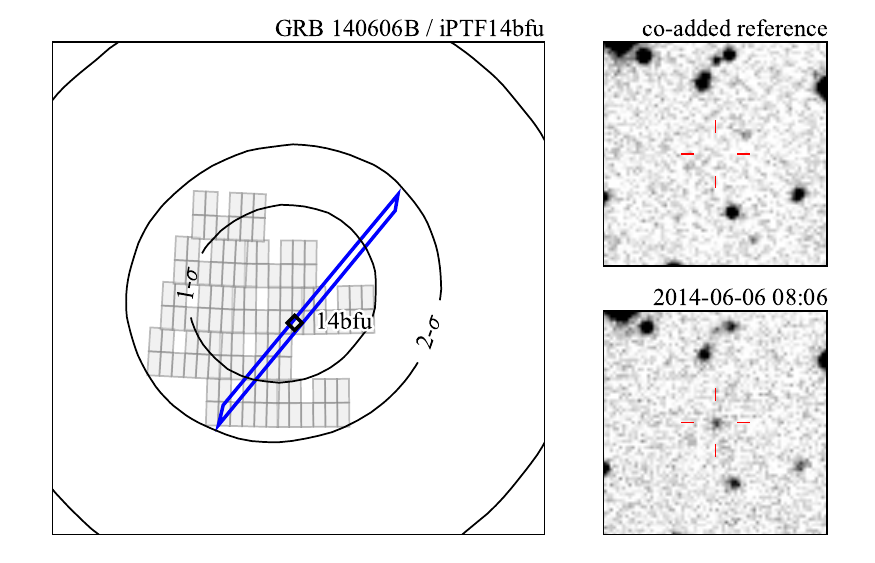}
\includegraphics{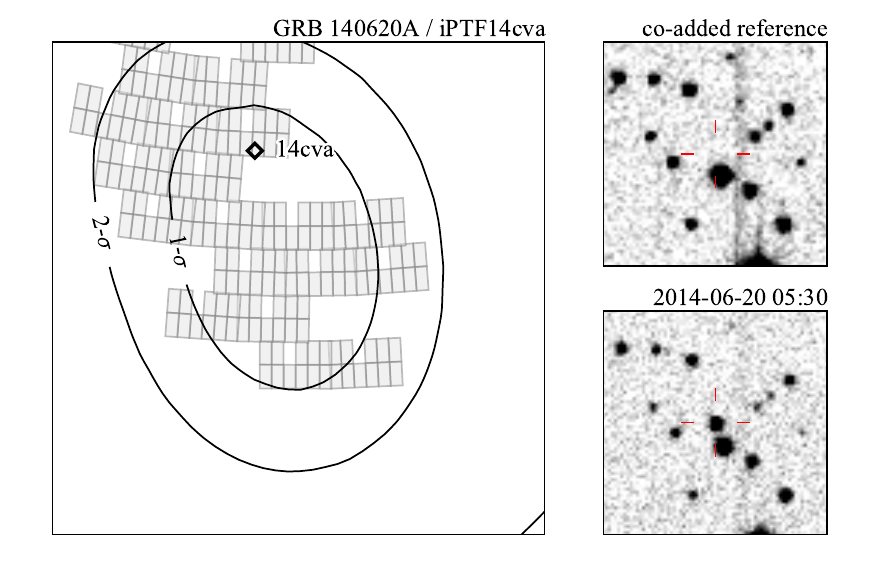}\\
\includegraphics{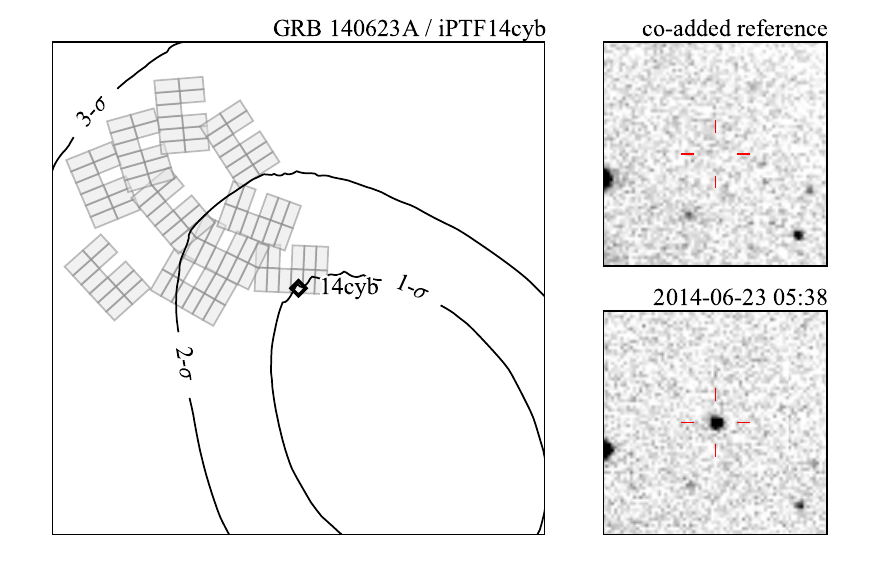}
\includegraphics{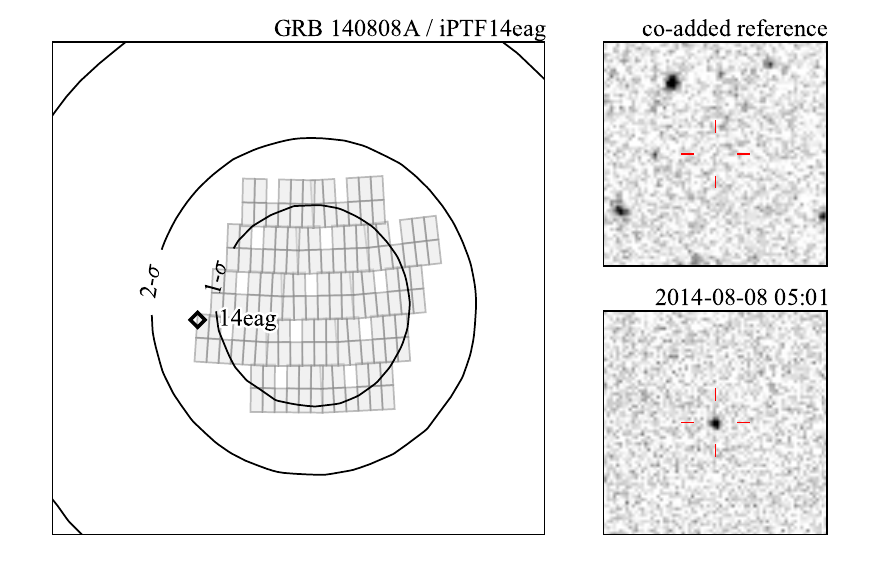}
\caption[Discovery images]{\label{fig:discovery-images}Gamma\nobreakdashes-ray localizations, \ac{P48} tiles, and discovery images for the \ac{GBM}\nobreakdashes--\ac{IPTF} afterglows. The \emph{Fermi} \ac{GBM} 1- and 2\nobreakdashes-$\sigma$ regions are shown as black contour lines, the \ac{P48} tiles as gray rectangles, the 3$\sigma$ \ac{IPN} triangulations in blue (when available), and the \ac{LAT} 1$\sigma$ error circles in green (when available). The positions of the optical transients are marked with black diamonds.}
\end{figure*}
\fi

\section{The Population in Context}

\subsection{Selection Effects}

First, we investigate the properties of the subset of \ac{GBM} bursts followed up by \ac{IPTF} compared to the \ac{GBM} bursts as a whole. It is known that, on average, GRBs with larger prompt fluences have brighter optical afterglows, though the correlation is very weak \citep{ComparisonAfterglows}. In Figure~\ref{fig:fluence-radius}, we plot the fluence in the 10\nobreakdashes--1000~keV band and 1\nobreakdashes-$\sigma$ localization radius of all \ac{GBM} bursts from the beginning of our experiment, retrieved from the \emph{Fermi} \ac{GBM} Burst Catalog at HEASARC.\footnote{\url{http://heasarc.gsfc.nasa.gov/W3Browse/fermi/fermigbrst.html}} As expected, there is a weak but clearly discernible correlation between fluence and radius, $F \propto r^{-1.3}$, with a Pearson correlation coefficient of $R = 0.64$.\footnote{In a separate sample of \ac{GBM} \acp{GRB} compiled by \citet{GBMLocalization}, the correlation between error radius and photon fluence is slightly stronger than the correlation between error radius and fluence. However, we use fluence rather than photon fluence here because the latter is not available for all bursts in the online \emph{Fermi} GBM archive.} The subset of bursts that we followed up spans a wide range in fluence and error radii up to $\sim10\arcdeg$. The bursts for which we detected optical afterglows are preferentially brighter, with the faintest burst having a fluence as low as $3\times10^{-6}$~erg\,cm$^{-2}$. There are some bright ($> 3 \times 10^{-5}$~erg\,cm$^{-2}$) and well\nobreakdashes-confined ($< 1.8\arcdeg$) events for which we did not find afterglows: those at 2013~August~28~07:19:56, 2013~November~27~14:12:14, and 2014~January~04~17:32:00 (see Table~\ref{table:nondetections}). However, these non\nobreakdashes-detections are not constraining given their ages of 20.28, 13.46, and 18.57 hours respectively. Conversely, there were two especially young bursts (followed up at $\Delta t = 0.11$~and~0.17\,hr) for which we did not detect afterglows. The non\nobreakdashes-detection of the burst at 2014~July~16~07:20:13 makes sense because we searched only 28\% of the \ac{GBM} localization. The non\nobreakdashes-detection on 2014~Paril~04~04:06:48, for which we observed 69\% of the localization, is a little more surprising, especially given its relatively high fluence of $8\times10^{-6}$~erg\,cm$^{-2}$; this is a possible candidate for a ``dark \ac{GRB}.'' On the whole, however, we can see that (1) we have followed up bursts with a large range of error radii and fluences, (2), there is a weak preference toward detecting bursts with small error radii, and (3) the detections tend toward bursts with high fluences. Naively one might expect higher fluences to translate into lower redshifts, but the interplay between the \ac{GRB} luminosity function and detector threshold greatly complicates such inferences \citep{GRBLuminosityFunction}.

\begin{figure}
    \centering
    \includegraphics{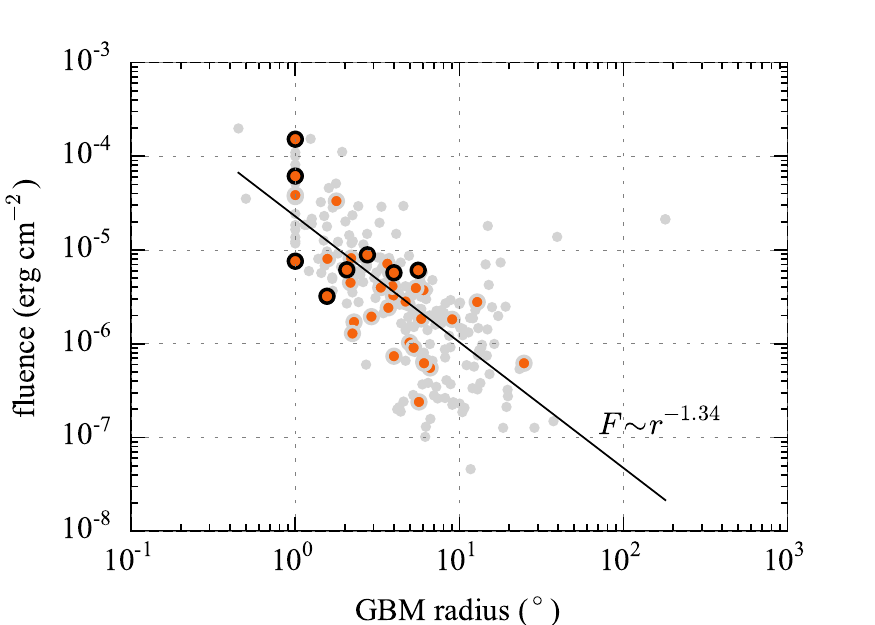}
    \caption[Fluence and error radius]{\label{fig:fluence-radius}Fluence and statistical error radius of \ac{GBM} bursts. Orange dots mark bursts that were followed up with \ac{IPTF}; black circles around orange dots mark bursts whose afterglows were detected by \ac{IPTF}. The black line is a power\nobreakdashes-law fit.}
\end{figure}

Second, the rich sample of all of the \ac{GRB} afterglows that we have today is undeniably the result of the success of the \emph{Swift} mission. It is therefore interesting to consider how the \ac{GBM}\nobreakdashes--\ac{IPTF} sample is similar to or different from the \emph{Swift} sample, given the differences in bandpasses and our increased reliance on the optical afterglow. In Figure~\ref{fig:redshift-distribution}, we plot the cumulative redshift distribution of our sample, alongside the distribution of redshifts of long \acp{GRB} detected by \emph{Swift}.\footnote{This sample was extracted from the \emph{Swift} GRB Table, \url{http://swift.gsfc.nasa.gov/archive/grb_table/}.} Indeed, we find that our sample is at lower redshifts; the former distribution lies almost entirely to the left of the latter, and the ratio of the median redshifts ($z = 1.5$ versus $z = 1.9$) of the two populations is about 0.75. However, with the small sample size, the difference between the two redshift distributions is not significant: a two\nobreakdashes-sample Kolmogorov\nobreakdashes--Smirnov test yields a $p$\nobreakdashes-value of 0.26, meaning that there is a 26\% chance of obtaining these two empirical samples from the same underlying distribution. More \ac{GBM}\nobreakdashes--\ac{IPTF} events are needed to determine whether the redshift distribution is significantly different.

\begin{figure}
    \centering
    \includegraphics{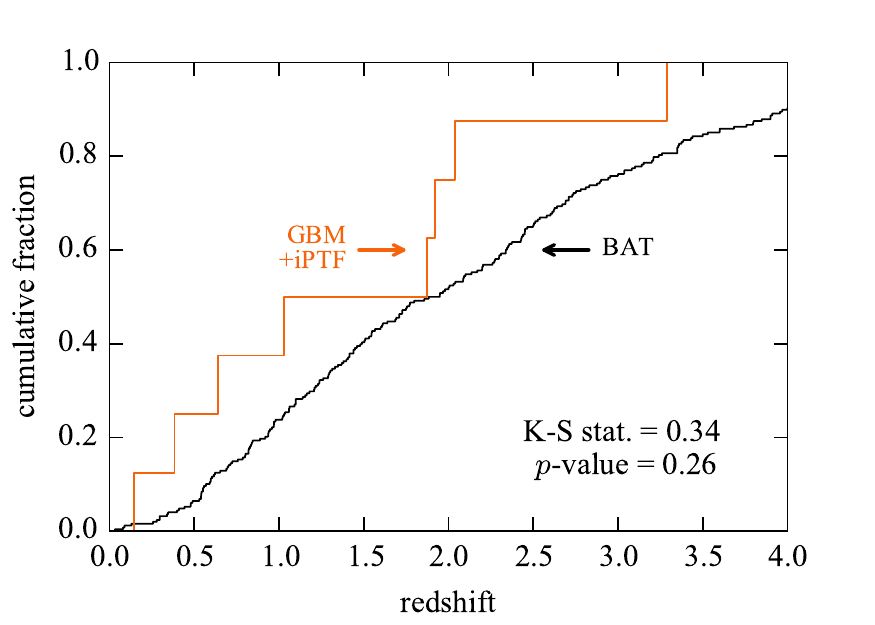}
    \caption[Cumulative redshift distribution]{\label{fig:redshift-distribution}Cumulative distribution of redshifts of long \acp{GRB} observed by \emph{Swift} \ac{BAT} (gray) and the \ac{GBM}\nobreakdashes--\ac{IPTF} experiment (orange).}
\end{figure}

\subsection{\acp{GRB} as Standard Candles?}

\begin{figure*}
    \centering
    \ifinthesis
\else
    \includegraphics{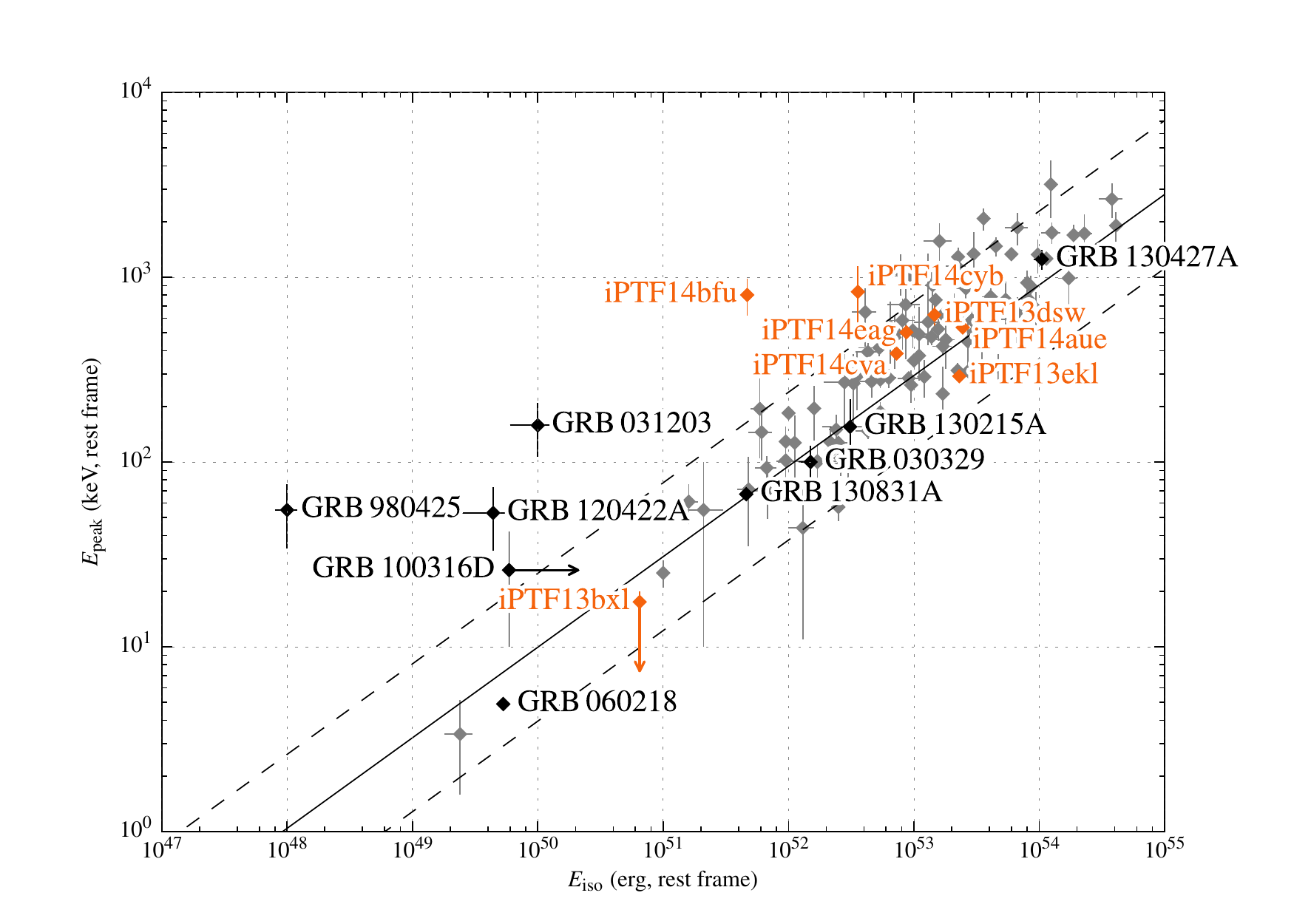}
    \fi
    \caption[Amati relation]{\label{fig:amati}Rest\nobreakdashes-frame energetics of \ac{GBM}\nobreakdashes--iPTF bursts (in orange) in comparison to an illustrative sample of previous \ac{GRB}\nobreakdashes--\acp{SN} (in black; includes \ac{GRB}\,060218/SN2006aj, \citealt{GRB060218-SN2006aj-1}, \citealt{GRB060218-SN2006aj-2}, \citealt{GRB060218-SN2006aj-3}; \ac{GRB}\,100316D/SN2010bh, \citealt{GRB100316D-SN2010bh-1}, \citealt{GRB100316D-SN2010bh-2}; \ac{GRB}\,120422A/SN2012bz, \citealt{GRB120422A-SN2012bz-2}, \citealt{GRB120422A-SN2012bz-1}; \ac{GRB}\,130215A/SN2013ez, \citealt{GRB130215A-SN2013ez}; \ac{GRB}\,130427A/SN2013cq, \citealt{GRB130427A-SN2013cq}; and \ac{GRB}\,130831A/SN2013fu, \citealt{GCN15320}). A general long \ac{GRB} sample from \citet{2006MNRAS.372..233A} and \citet{2008MNRAS.391..577A,2009A&A...508..173A} is shown in gray. The solid black line represents the Amati relation as given in \citet{2006MNRAS.372..233A}, $E_\mathrm{peak} = 95 (E_{\gamma,\mathrm{iso}} / 10^{52}\text{ erg})^{0.49}$~keV. The black dashed lines show the relation's $1\sigma$ dispersion of $\pm0.4$~dex.}
\end{figure*}

\citet{AmatiRelation} pointed out a striking empirical correlation in the rest\nobreakdashes-frame prompt emission spectra of BeppoSAX \acp{GRB}, with the peak energy (in the $\nu F_\nu$ sense) $E_\mathrm{peak}$ related to the bolometric, isotropic\nobreakdashes-equivalent energy release $E_\mathrm{iso}$ by $E_\mathrm{peak} \propto {E_\mathrm{iso}}^m$. It was quickly realized that such a relation, if intrinsic to the bursts, could be used to measure the redshifts of \acp{GRB} non\nobreakdashes-spectroscopically \citep{EmpiricalRedshiftIndicators}. As with the Phillips relation for \acp{SN}~Ia \citep{PhillipsRelation}, with such a relation \acp{GRB} could serve as standardizable candles in order to measure cosmological parameters (\citealt{DaiGRBsStandardCandles,FriedmanGRBsStandardCandles,GhirlandaGRBsStandardCandles}; etc.).

However, there has been a vigorous debate about whether the Amati relation and related correlations are innate to \acp{GRB} or reflect a detector\nobreakdashes-dependent selection bias \citep{2005ApJ...627..319B,2005MNRAS.361L..10G,2005MNRAS.360L..73N,2006ApJ...636L..73S,2007ApJ...671..656B,2007MNRAS.382..342C,2007ApJ...656L..53S,2009ApJ...694...76B,2009MNRAS.393.1209F,2009ApJ...704.1405K,GRBLuminosityFunction,2011MNRAS.411.1843S,2012ApJ...747...39C,2012ApJ...747..146K}. One alternative interpretation is that bursts to the upper\nobreakdashes-left boundary of the Amati relation are selected against by photon\nobreakdashes-counting instruments because, being relatively hard, there are fewer photons. The lack of bursts to the lower right of the Amati line may be due to a genuine lack of relativistic explosions that are much softer than, but as energetic as, standard \acp{GRB}.

It has been difficult to directly test the Amati relation in the context of \emph{Fermi} bursts because most lack known redshifts, since bursts that were coincidentally observed and localized by the \emph{Swift} \ac{BAT} do not directly sample the selection bias of \emph{Fermi} \ac{GBM}. However, \citet{FermiAmatiDebate} showed that many \emph{Fermi} bursts that lack known redshifts would be inconsistent with the Amati relation at any distance. (See also \citealt{UrataEnergeticWAMLATGRBs} for outlier events detected by \emph{Fermi}~\ac{LAT} and \emph{Suzaku}~\acs{WAM}.) Here, we have a small sample of \emph{Fermi} bursts with \emph{known} redshifts. One of them, \ac{GRB}\,140606B/iPTF14bfu at $z=0.384$, is a clear outlier, over $2\sigma$ away from the mean Amati relation. This burst is not alone: in Figure~\ref{fig:amati}, we have marked a selection of previous long \acp{GRB} with spectroscopically identified \acp{SN}. Three among them are also outliers. (A possible caveat is that the prompt emission mechanism for \ac{GRB}\,140606B could be different from typical cosmological bursts; we explore this in the next section.) To be sure, most of the bursts in our \ac{GBM}\nobreakdashes--\ac{IPTF} sample fall within a $1\sigma$ band of the Amati relation. This includes the nearest event to date, \ac{GRB}\,130702A/iPTF13bxl at $z=0.145$. However, the one outlier in our admittedly small sample strengthens the case that the boundary of the Amati relation is somewhat influenced by the detector thresholds and bandpasses.

\subsection{Shock Breakout}

Two \acp{GRB} in our sample, \ac{GRB}\,130702A/iPTF13bxl and \ac{GRB}\,140606B/iPTF14bfu, have $E_\mathrm{iso} \sim 10^{51}$~erg (rest frame), energetically intermediate between ``standard'' luminous, cosmically distant bursts and nearby \acp{llGRB}. Prototypes of the latter class include \ac{GRB}\,980425/\ac{SN}~1998bw \citep{SN1998bw,1998bwShockBreakout}, which was also the first \ac{SN} discovered in association with a \ac{GRB}. They offer an interesting test case for competing theories to explain the wide range of prompt gamma\nobreakdashes-ray energy releases observed from \acp{GRB} (e.g., \citealt{2014A&A...566A.102S}).

It has been suggested that the two luminosity regimes correspond to different prompt emission mechanisms \citep{AreLowLuminosityBurstsGeneratedByRelativisticJets}. The \acp{llGRB} could be explained by the breakout of a mildly relativistic shock from the progenitor envelope \citep{RelativisticShockBreakoutRelation}. High\nobreakdashes-luminosity bursts, on the other hand, are thought to be produced by internal shocks within an ultra\nobreakdashes-relativistic jet \citep{ReesInternalShocks} that has successfully punched through the star. A central engine that sometimes fails to launch an ultra\nobreakdashes-relativistic jet is one way to unify the luminosity functions of standard \acp{GRB} and \acp{llGRB} \citep{LuminosityFunctionJetStructureGRBs}.

The smoking gun for the relativistic shock breakout model is a cooling, thermal component to the prompt X\nobreakdashes-ray emission, as in the case of \ac{GRB}\,060218 \citep{GRB060218ShockBreakoutNature}. Unfortunately, this diagnostic is not possible for \ac{GRB}\,130702A and \ac{GRB}\,140606B because we lack early\nobreakdashes-time \emph{Swift} observations.

However, \citet{RelativisticShockBreakoutRelation} propose a closure relation (their Equation (18)) between the prompt energy, temperature, and timescale that is valid for shock breakout\nobreakdashes-powered \acp{GRB}. We reproduce it here:
\begin{equation}
    t^\mathrm{obs}_\mathrm{bo} \sim
        20~\text{s}
        \left(\frac{E_\mathrm{bo}}{10^{46}~\text{erg}}\right)^{\frac{1}{2}}
        \left(\frac{T_\mathrm{bo}}{50~\text{keV}}\right)^{-\frac{9+\sqrt{3}}{4}}.
\end{equation}
If we very crudely assume that all of the prompt emission is from a shock escaping from the progenitor envelope, then we can use $E_\mathrm{iso}$, $E_\mathrm{peak}$, and $T_{90}$ as proxies for those observables. This gives us a simple discriminator of which bursts are plausible shock breakout candidates, the ratio
\begin{equation}
    \xi = (1 + z) t^\mathrm{obs}_\mathrm{bo} / T_\mathrm{90},
\end{equation}
which should be close to 1. As expected, most of the energetic ($E_\mathrm{iso} > 10^{52}~\textrm{erg}$), cosmic ($z > 0.5$) \acp{GRB} in our sample are inconsistent with the closure relation. They are all much \emph{shorter} in duration, given their $\gamma$\nobreakdashes-ray spectra, than would be expected for a shock breakout. The exception is \ac{GRB}\,140623A/iPTF14cyb, which yields $\xi = 0.5 \pm 0.5$. In this particular case, one possible explanation is that the central engine simply remained active for much longer than the timescale of the shock breakout.

Surprisingly, of the two low\nobreakdashes-luminosity, low\nobreakdashes-redshift bursts in our sample, \ac{GRB}\,130702A/iPTF13bxl's prompt emission was much too brief to be consistent with this shock breakout model, with $\xi = (1.6 \pm 0.7) \times 10^3$. Most likely, this means that the prompt emission of \ac{GRB}\,130702A is simply a very soft, very sub\nobreakdashes-luminous version of an otherwise `ordinary' long \ac{GRB}. Any early\nobreakdashes-time shock breakout signature, if present, was not observed either because it occurred at energies below \ac{GBM}'s bandpass or because it was much weaker than the emission from the standard \ac{GRB} mechanism. However, \ac{GRB}\,140606B/iPTF14bfu's prompt emission is consistent with the closure relation, with $\xi = 0.5 \pm 0.3$. Though we must interpret this with caution because we cannot disentangle a thermal component from the \ac{GBM} data, if we naively apply linear least squares to (the logarithm of) Equations (14), (16), (17) of \citet{RelativisticShockBreakoutRelation},
\begin{align}
    E_\mathrm{bo} &\approx 2 \times 10^{45} R_5^2 \gamma_{f,0}^\frac{1+\sqrt{3}}{2}~\text{erg}, \\
    T_\mathrm{bo} &\sim 50 \gamma_{f,0}~\text{keV}, \\
    t_\mathrm{bo}^\mathrm{obs} &\approx 10 \frac{R_5}{\gamma_{f,0}^2}~\text{s},
\end{align}
then we find the breakout radius and Lorentz factor to be
\begin{eqnarray*}
    R_\mathrm{bo} &=& (1.3 \pm 0.2) \times 10^3 \, R_\sun, \\
    \gamma_{f,0} &=& 14 \pm 2.
\end{eqnarray*}

The breakout radius is comparable to that which \citet{RelativisticShockBreakoutRelation} find for \ac{GRB}\,060218 and \ac{GRB}\,100316D, suggestive of breakout from a dense wind environment, rather than the star itself.\footnote{Note that SN~2008D, which seems to be the only case so far of shock breakout observed in an ``ordinary'' \ac{SN}~Ibc, had a 500~s emission episode that was not strictly consistent with the picture of shock breakout from a progenitor envelope.  \citet{SN2008DShockBreakoutThickWind,SpectrumLightCurveShockBreakoutThickWind} explore the case of shock breakout through a thick Wolf\nobreakdashes--Rayet wind, which can accommodate longer emission timescales.} However, the derived Lorentz factor of \ac{GRB}\,140606B is a bit higher than those of the other two examples.

Another way to constrain the nature of the explosion is to look at the kinetic energy $E_{k,\mathrm{iso}}$ of the blast compared to the promptly radiated energy $E_{\gamma,\mathrm{iso}} \equiv E_\mathrm{iso}$ and the radiative efficiency $\eta = E_{\gamma,\mathrm{iso}} / (E_{k,\mathrm{iso}} + E_\mathrm{iso})$. After the end of any plateau phase, the X\nobreakdashes-ray flux is a fairly clean diagnostic of $E_{k,\mathrm{iso}}$ assuming that the X\nobreakdashes-rays are above the cooling frequency \citep{EnergyOfGammaRayBursts}. During the slow\nobreakdashes-cooling phase and under the typical conditions where $p \approx 2$ and $\nu_\mathrm{c} < \nu_\mathrm{X}$, the X\nobreakdashes-ray flux is only weakly sensitive to global parameters such as the fraction of the internal energy partitioned to electrons and to the magnetic ($\epsilon_e$, $\epsilon_B$). Even the radiative losses, necessary for extrapolating from the late\nobreakdashes-time afterglow to the end of the prompt phase, are minor, amounting to order unity at $\Delta t = 1$~day \citep{KineticEnergyRadiativeEfficiencyOfGammaRayBursts}. We calculate the isotropic-equivalent rest\nobreakdashes-frame X\nobreakdashes-ray luminosity from the flux at $\Delta t = 1$~day using Equation~(1) of \citet{FermiSwiftPopulationStudies}, reproduced below:
\begin{equation}\label{eq:LX}
    L_\mathrm{X}(t) = 4 \pi {D_\mathrm{L}}^2 F_\mathrm{X} (t) (1 + z)^{-\alpha_\mathrm{X} + \beta_\mathrm{X} - 1}.
\end{equation}
Then we estimate the kinetic energy at the end of the prompt emission phase using Equation~(7) of \citet{KineticEnergyRadiativeEfficiencyOfGammaRayBursts}:
\begin{multline}\label{eq:Ek}
    E_{k,\mathrm{iso}} = \left(10^{52}~\text{ergs}\right) \times R \times
        \left(\frac{L_\mathrm{X} (1~\text{day})}{10^{46}~\text{ergs}~\text{s}^{-1}}\right)^{-4/(p+2)}
        \left(\frac{1 + z}{2}\right)^{-1} \\
        \times \epsilon_{e,-1}^{4(1-p)/(2+p)}
        \epsilon_{B,-2}^{(2-p)/(2+p)} t_\mathrm{10~hr}^{(3p-2)/(p+2)}
        \nu_{18}^{2(p-2)/(p+2)}.
\end{multline}
The correction factor $R$ for radiative losses is given by Equation~(8) of \citet{KineticEnergyRadiativeEfficiencyOfGammaRayBursts}, adopted here:
\begin{equation}\label{eq:R}
    R = \left(\frac{t}{T_\mathrm{90}}\right)^{(17/16)\epsilon_e}.
\end{equation}
The numeric subscripts follow the usual convention for representing quantities in powers of 10 times the cgs unit, i.e., $\epsilon_{e,-1} = \epsilon_e / 10^{-1}$, $\epsilon_{B,-2} = \epsilon_B / 10^{-2}$, and $\nu_{18} \equiv \nu / (10^{18}~\text{Hz})$. We assume $\epsilon_e = 0.1$ and $\epsilon_B = 0.01$. For bursts that have \ac{XRT} detections around $\Delta t = 1$~day (\acp{GRB}\,130702A, 131231A, 140508A, 140606B, and 140620A), we calculate $L_\mathrm{X}$ by interpolating a least\nobreakdashes-squares power\nobreakdashes-law fit to the X\nobreakdashes-ray light curve. Some of our bursts (\acp{GRB}\,131011A, 140623A, and 140808A) were only weakly detected by \ac{XRT}; for these we extrapolate from the mean time of the \ac{XRT} detection assuming a typical temporal slope of $\alpha_\mathrm{X} = 1.43 \pm 0.35$ \citep{FermiSwiftPopulationStudies}. The kinetic and radiative energies of our eight bursts are shown in Figure~\ref{fig:radiative-efficiency}. Half of our bursts are reasonably well constrained in $E_\mathrm{k}$--$E_\gamma$ space; these are shown as red points. The other half (\acp{GRB}\,131011A, 131231A, 140620A, and 140623A) have highly degenerate \acp{SED}, so their position in this plot is highly sensitive to model assumptions; these are shown as gray points. Dotted lines are lines of constant radiative efficiency.

Within our sample, there are at least three orders of magnitude of variation in both $E_{k,\mathrm{iso}}$ and $E_{\gamma,\mathrm{iso}}$. The two \ac{GRB}\nobreakdashes--\acp{SN} have radiative and kinetic energies of $\sim 10^{51}$~erg, both two to three orders of magnitude lower than the other extreme in our sample or the average values for \emph{Swift} bursts. In our sample, they have two of the lowest inferred radiative efficiencies of $\eta \sim 0.1$\nobreakdashes--0.5, but these values are not atypical of BATSE bursts (e.g., \citealt{KineticEnergyRadiativeEfficiencyOfGammaRayBursts}) and are close to the median value for \emph{Swift} bursts. These are, therefore, truly less energetic than cosmological bursts, not merely less efficient at producing gamma rays.

\begin{figure}
    \centering
    \includegraphics{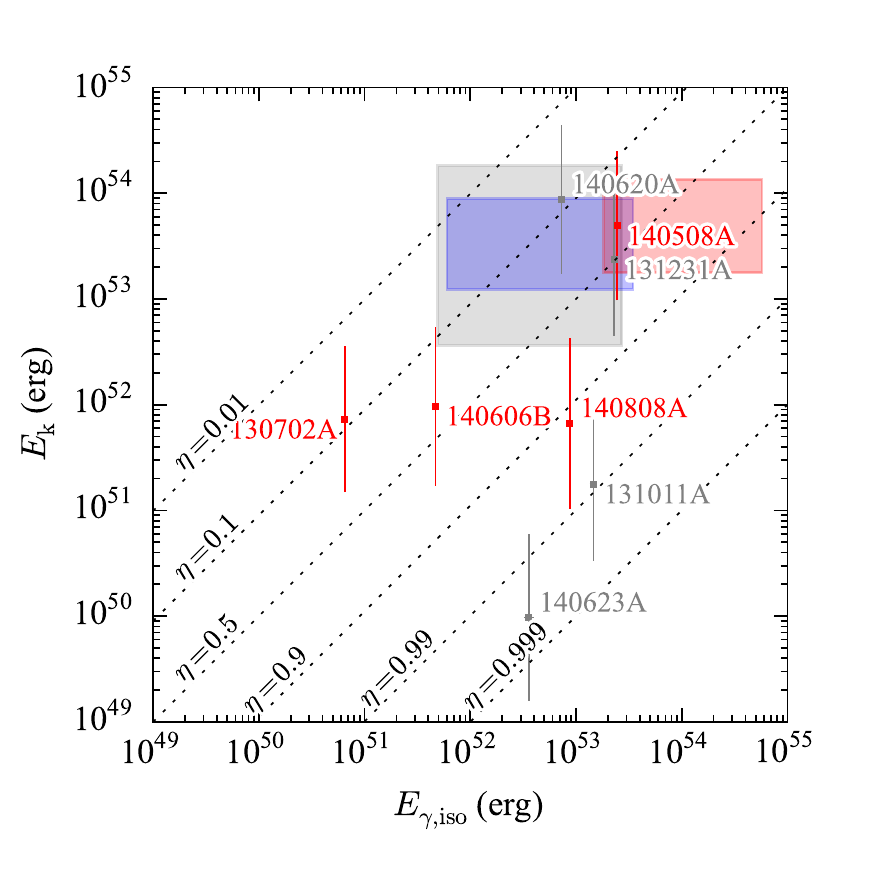}
    \caption[Radiative efficiency of prompt phase]{\label{fig:radiative-efficiency}Fireball kinetic energy $E_{k,\mathrm{iso}}$ at $t = T_{90}$ as estimated from X\nobreakdashes-ray flux vs. rest\nobreakdashes-frame isotropic\nobreakdashes-equivalent gamma\nobreakdashes-ray energy $E_{\gamma,\mathrm{iso}}$. Red points denote bursts for which $E_{k,\mathrm{iso}}$ can be reliably estimated from the \emph{Swift} \ac{XRT} data; gray points denote bursts for which the calculation of $E_{k,\mathrm{iso}}$ may have extreme model dependence. Dashed lines are lines of constant radiative efficiency $\eta = E_{\gamma,\mathrm{iso}} / (E_{k,\mathrm{iso}} + E_{\gamma,\mathrm{iso}})$. The gray, blue, and red rectangles show the 1$\sigma$ parameter ranges of \emph{Swift} \ac{BAT}, \ac{BAT}+\ac{GBM}, and \ac{BAT}+\ac{LAT} long \acp{GRB} from \citet{FermiSwiftPopulationStudies}.}
\end{figure}

\section{Looking Forward}

In this experiment, we have followed up 35 \emph{Fermi} \ac{GBM} bursts, scanning areas from 30~to~147~deg$^2$. To date, we have detected eight afterglows with apparent optical magnitudes as bright as $R \approx 16$ and as faint as $R \approx 20$. We have found redshifts as nearby as $z = 0.145$ and as distant as $z = 3.29$. A continuation of the project should reveal more low\nobreakdashes-redshift events, more GRB\nobreakdashes--\acp{SN}, and more relatively hard \acp{GRB}.

We aim to uncover the much fainter afterglows of short, hard bursts by using stacked \ac{P48} exposures and integrating a co-addition stage into the real-time pipeline, and by honing our follow\nobreakdashes-up to sift through the increased number of candidates. The greatest factor limiting discoveries is, of course, that \emph{Fermi} detects bursts all over the sky, only a fraction of which are visible from Palomar. Given our success so far, we enthusiastically suggest that other wide\nobreakdashes-field surveys implement a similar program. Furthermore, automatically sharing lists of candidates between longitudinally separated instruments would facilitate rapid identification and follow\nobreakdashes-up of the fastest\nobreakdashes-fading events.

It is uncertain what directions future gamma\nobreakdashes-ray space missions will take. Some may be like \emph{Swift}, able to rapidly train multiple on\nobreakdashes-board follow\nobreakdashes-up instruments on new targets. Even if they lack these capabilities, we should be able to routinely locate \ac{GRB} afterglows and find their redshifts using targeted, ground\nobreakdashes-based optical transient searches similar to the one that we have described.

Looking beyond \acp{GRB}, our present effort serves as a prototype for searching for optical counterparts of \ac{GW} transients. We expect that many of the techniques that we have described and the lessons that we have learned in the context of \ac{IPTF} will generalize to other wide-field instruments on meter-class and larger telescopes.

Near the end of 2015, Advanced \ac{LIGO} will begin taking data, with Advanced Virgo soon following suit. The first binary neutron star merger detections are anticipated by 2016 or later \citep{LIGOObservingScenarios}. On a similar timescale, \ac{IPTF} will transform into the \acl{ZTF}, featuring a new 47~deg$^2$ survey camera that can reach $R = 20.4$~mag~in~30~s. The prime \ac{GW} sources, \ac{BNS} mergers, may also produce a variety of optical transients: on- or off\nobreakdashes-axis afterglows \citep{SyntheticSGRBAfterglows,UrataOffAxisSoftXRF}, kilonovae \citep{Kilonova,KilonovaHighOpacities}, and neutron-powered precursors \citep{KilonovaPrecursor}; see Figure~\ref{fig:sgrb-kcorrected} for some examples.

There will be two key challenges. First, \ac{GW} localizations can be even coarser than \emph{Fermi} \ac{GBM} error circles. Starting around $\sim$600~deg$^2$ in the initial (2015) two\nobreakdashes-detector configuration \citep{KasliwalTwoDetectors,FirstTwoYears}, the areas will shrink to $\sim$200~deg$^2$ with the addition of Virgo in 2016. They should reach $\sim 10$~deg$^2$ toward the end of the decade as the three detectors approach final design sensitivity and can approach $\sim 1$~deg$^2$ as additional planned \ac{GW} facilities come online (LIGO\nobreakdashes--India and KAGRA; see \citealt{ShutzThreeFiguresOfMerit,Veitch:2012,FairhurstLIGOIndia,NissankeKasliwalEMCounterparts,LIGOObservingScenarios}). Since the detection efficiency of our \ac{GBM}--\ac{IPTF} afterglow search is consistent with the areas that we searched, we expect that even the earliest Advanced \ac{LIGO} localizations will present no undue difficulties for \ac{ZTF} when we consider its 15\nobreakdashes-fold increase in areal survey rate as compared to \ac{IPTF}.

However, there is a second challenge that these optical signatures are predicted to be fainter than perhaps $22$~mag (with the exception of on-axis afterglows, which should be rare but bright due to beaming). For meter-size telescopes, this will require integrating for much longer (10~minutes to 1~hr) than we have been performing with \ac{IPTF}. Fortunately, because the \ac{LIGO} antenna pattern is preferentially sensitive above and directly opposite of North America, we are optimistic that many early Advanced \ac{LIGO} events should be promptly accessible from Palomar with long observability windows \citep{KasliwalTwoDetectors}.

The main difficulty for any \ac{GW} optical counterpart search will be the inundation of false positives due to the required depth and area. We enumerate the following strategies to help identify the one needle in the haystack:
\begin{enumerate}
\item Improved machine learning algorithms (see \citealt{RB4} in the context of \ac{IPTF}) will decrease the contamination of the discovery stream by artifacts.
\item Combining a catalog of nearby galaxies with the distance and position information from the \ac{GW} observations can help to reduce and prioritize targets for further follow\nobreakdashes-up \citep{NissankeKasliwalEMCounterparts}.
\item Better leveraging of light\nobreakdashes-curve history across multiple surveys will help to automate the selection of targets for photometric follow-up with multiple telescopes.
\item Our first experiences with detections and non-detections will guide decisions about the optimal filter. At the moment, kilonova models prefer redder filters (suggesting $i$ band), and precursor models prefer bluer (suggesting $g$ band).
\end{enumerate}

The combination of gamma-ray missions, ground-based \ac{GW} detectors, and synoptic optical survey instruments is poised to make major discoveries over the next few years, of which we have provided a small taste in this work. We offer both lessons learned and a way forward in this multimessenger effort. The ultimate reward will be joint observations of a compact binary merger in gamma, X\nobreakdashes-rays, optical, and GWs, giving us an exceptionally complete record of a complex astrophysical process: it will be almost as good as being there.
 
\makeatletter{}\begin{figure}
    \centering
    \includegraphics{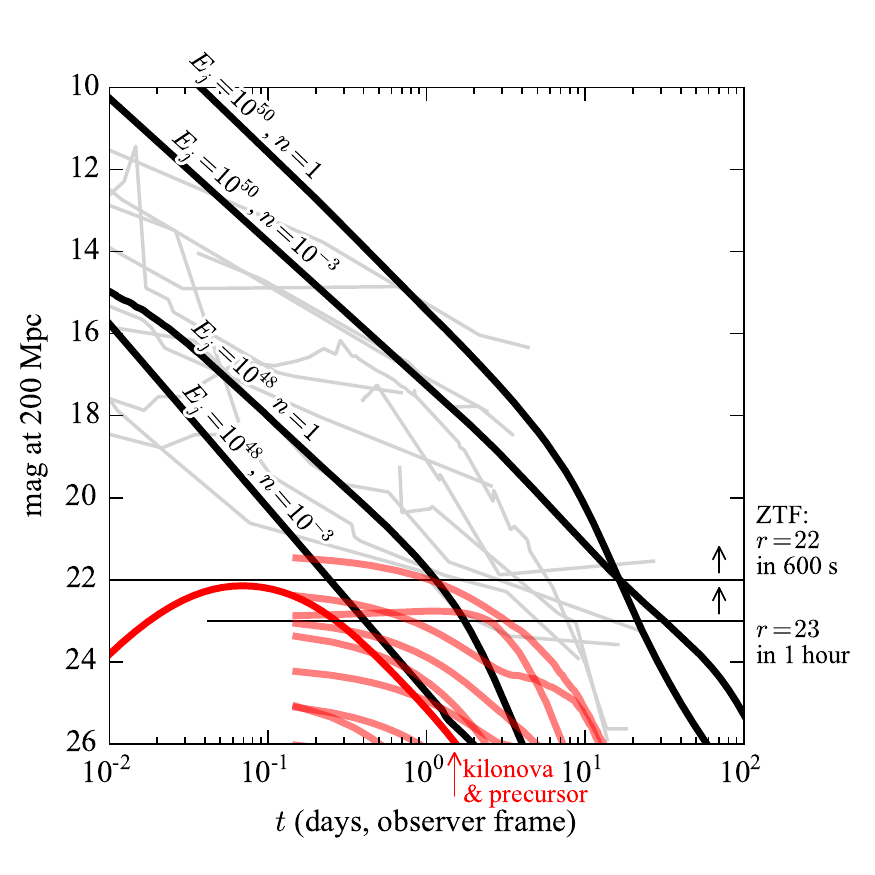}
    \caption[Typical light curves of Advanced \acs{LIGO} optical counterparts]{\label{fig:sgrb-kcorrected}Light curves of short \ac{GRB} afterglows, scaled to an Advanced \ac{LIGO} range of 200~Mpc. Thin gray lines are afterglows of \emph{Swift} short \acp{GRB} that have known redshifts. Thick black lines are synthetic on\nobreakdashes-axis afterglows from \citet{SyntheticSGRBAfterglows} with jet half\nobreakdashes-opening angles of 0.2~rad and observer angles of 0~rad. Jet energies $E_j$, in units of erg, and circumburst densities $n$, in units of cm$^{-3}$, are labeled on the plot. The solid, deep\nobreakdashes-red line is the $r$\nobreakdashes-band neutron\nobreakdashes-powered kilonova precursor model from \citet{KilonovaPrecursor} with opacity $\kappa_r = 30$~cm$^2$\,g$^{-1}$, free neutron mass $m_n = 10^{-4}$~$M_\sun$, and electron fraction $Y_e = 0.05$. The solid, light\nobreakdashes-red lines represent kilonova models from \citet{KilonovaHighOpacities} with ejected masses of $m_\mathrm{ej} = 10^{-3}$,~$10^{-2}$,~or $10^{-1}~M_\sun$ and characteristic velocities $\beta = v/c = 0.1$,~0.2, or~0.3. The \citet{KilonovaPrecursor} kilonova precursors are blue: they peak at about 0.1~mag brighter in the $g$ band than in $r$. The \citet{KilonovaHighOpacities} kilonova models are red: they are about 1~mag brighter in the $i$ band than in $r$.}
\end{figure}

\acknowledgements

\makeatletter{}L.P.S. thanks generous support from the \ac{NSF} in the form of a Graduate Research Fellowship.

The National Radio Astronomy Observatory is a facility of the \ac{NSF} operated under cooperative agreement by Associated Universities, Inc.

This paper is based on observations obtained with the \acl{P48} and the \acl{P60} at the Palomar Observatory as part of the Intermediate Palomar Transient Factory project, a scientific collaboration among the California Institute of Technology, Los Alamos National Laboratory, the University of Wisconsin, Milwaukee, the Oskar Klein Center, the Weizmann Institute of Science, the TANGO Program of the University System of Taiwan, and the Kavli Institute for the Physics and Mathematics of the Universe. The present work is partly funded by \emph{Swift} Guest Investigator Program Cycle 9 award 10522 (NASA grant NNX14AC24G) and Cycle 10 award 10553 (NASA grant NNX14AI99G).

Some of the data presented herein were obtained at the W. M. Keck Observatory, which is operated as a scientific partnership among the California Institute of Technology, the University of California, and NASA; the Observatory was made possible by the generous financial support of the W.~M.~Keck Foundation.

We thank Thomas Kr\"{u}hler for reducing the X-shooter spectrum of GRB~131011A~/~iPTF13dsw.

We thank the staff of the Mullard Radio Astronomy Observatory for their invaluable assistance in the operation of \ac{AMI}. G.E.A., R.P.F., and T.D.S. acknowledge the support of the European Research Council Advanced Grant 267697, ``4 Pi Sky: Extreme Astrophysics with Revolutionary Radio Telescopes.''

Support for \ac{CARMA} construction was derived from the Gordon and Betty Moore Foundation, the Kenneth T. and Eileen L. Norris Foundation, the James S. McDonnell Foundation, the Associates of the California Institute of Technology, the University of Chicago, the states of California, Illinois, and Maryland, and the \ac{NSF}. Ongoing \ac{CARMA} development and operations are supported by the \ac{NSF} under a cooperative agreement and by the \ac{CARMA} partner universities.

These results made use of Lowell Observatory's \ac{DCT}. Lowell operates the \ac{DCT} in partnership with Boston University, Northern Arizona University, the University of Maryland, and the University of Toledo. Partial support of the \ac{DCT} was provided by Discovery Communications. \ac{LMI} was built by Lowell Observatory using funds from the \ac{NSF} (AST-1005313).

This work is partly based on observations made with \ac{GTC}, at the Roque de los Muchachos Observatory (La Palma, Spain). The research activity of A.de.U.P., C.T., and J.G. is supported by Spanish research project AYA2012\nobreakdashes-39362\nobreakdashes-C02\nobreakdashes-02. A.d.U.P. acknowledges support by the European Commission under the Marie Curie Career Integration Grant programme (FP7\nobreakdashes-PEOPLE\nobreakdashes-2012\nobreakdashes-CIG~322307).

A portion of this work was carried out at the Jet Propulsion Laboratory under a Research and Technology Development Grant, under contract with NASA. Copyright 2014 California Institute of Technology. All Rights Reserved. US Government Support Acknowledged.

K.H. acknowledges support for the \ac{IPN} under the following NASA grants: NNX07AR71G, NNX13AP09G, NNX11AP96G, and NNX13AI54G.

The Konus\nobreakdashes-\emph{Wind} experiment is partially supported by a Russian Space Agency contract and RFBR grants 15\nobreakdashes-02\nobreakdashes-00532 and 13\nobreakdashes-02\nobreakdashes-12017\nobreakdashes-ofi\nobreakdashes-m.

IRAF is distributed by the National Optical Astronomy Observatory, which is operated by the Association of Universities for Research in Astronomy (AURA) under cooperative agreement with the \ac{NSF}.

This research has made use of data, software, and/or web tools obtained from \ac{HEASARC}, a service of the Astrophysics Science Division at NASA/GSFC and of the Smithsonian Astrophysical Observatory's High Energy Astrophysics Division.

This research has made use of \ac{NED}, which is operated by the Jet Propulsion Laboratory, California Institute of Technology, under contract with NASA.

This work made use of data supplied by the UK Swift Science Data Centre at the University of Leicester including the \emph{Swift} \ac{XRT} \ac{GRB} catalog and light\nobreakdashes-curve repository \citep{SwiftXRTRepository,2009MNRAS.397.1177E,2007A&A...476.1401G}.

This research made use of Astropy\footnote{\url{http://www.astropy.org}} \citep{astropy}, a community-developed core Python package for Astronomy. Some of the results in this paper have been derived using HEALPix \citep{HEALPix}.

{\it Facilities:} \facility{\emph{Fermi} (GBM, LAT)}, \facility{PO:1.2m (CFH12k)}, \facility{PO:1.5m}, \facility{Hale (DBSP)}, \facility{Gemini:Gillett (GMOS)}, \facility{Gemini:South (GMOS)}, \facility{EVLA}, \facility{CARMA}, \facility{\emph{Swift} (XRT, UVOT)}, \facility{Keck:I (LRIS)}, \facility{Keck:II (DEIMOS)}, \facility{NOT (ALFOSC)}, \facility{HCT}, \facility{AMI}, \facility{VLT (X-shooter)}, \facility{\emph{INTEGRAL} (SPI-ACS)}, \facility{Mars Odyssey (HEND)}, \facility{MESSENGER (GRNS)}, \facility{WIND (Konus)}

\appendix

We illustrate three stages of the \ac{IPTF} pipeline that we discussed in Section~\ref{sec:afterglow-search-method}: the \ac{TOO} Marshal (Figure~\ref{fig:screenshot}), the Treasures Portal (Figure~\ref{fig:treasures}), and the Transient Marshal (Figure~\ref{fig:transient-marshal}).

\makeatletter{}\providecommand{\acrolowercase}[1]{\lowercase{#1}}

\begin{acronym}
\acro{2MASS}[2MASS]{Two Micron All Sky Survey}
\acro{AdVirgo}[AdVirgo]{Advanced Virgo}
\acro{AMI}[AMI]{Arcminute Microkelvin Imager}
\acro{AGN}[AGN]{active galactic nucleus}
\acroplural{AGN}[AGN\acrolowercase{s}]{active galactic nuclei}
\acro{aLIGO}[aLIGO]{Advanced \acs{LIGO}}
\acro{ATCA}[ATCA]{Australia Telescope Compact Array}
\acro{ATLAS}[ATLAS]{Asteroid Terrestrial-impact Last Alert System}
\acro{BAT}[BAT]{Burst Alert Telescope\acroextra{ (instrument on \emph{Swift})}}
\acro{BATSE}[BATSE]{Burst and Transient Source Experiment\acroextra{ (instrument on \acs{CGRO})}}
\acro{BAYESTAR}[BAYESTAR]{BAYESian TriAngulation and Rapid localization}
\acro{BBH}[BBH]{binary black hole}
\acro{BHBH}[BHBH]{\acl{BH}\nobreakdashes--\acl{BH}}
\acro{BH}[BH]{black hole}
\acro{BNS}[BNS]{binary neutron star}
\acro{CARMA}[CARMA]{Combined Array for Research in Millimeter\nobreakdashes-wave Astronomy}
\acro{CASA}[CASA]{Common Astronomy Software Applications}
\acro{CFH12k}[CFH12k]{Canada--France--Hawaii $12\,288 \times 8\,192$ pixel CCD mosaic\acroextra{ (instrument formerly on the Canada--France--Hawaii Telescope, now on the \ac{P48})}}
\acro{CRTS}[CRTS]{Catalina Real-time Transient Survey}
\acro{CTIO}[CTIO]{Cerro Tololo Inter-American Observatory}
\acro{CBC}[CBC]{compact binary coalescence}
\acro{CCD}[CCD]{charge coupled device}
\acro{CDF}[CDF]{cumulative distribution function}
\acro{CGRO}[CGRO]{Compton Gamma Ray Observatory}
\acro{CMB}[CMB]{cosmic microwave background}
\acro{CRLB}[CRLB]{Cram\'{e}r\nobreakdashes--Rao lower bound}
\acro{cWB}[\acrolowercase{c}WB]{Coherent WaveBurst}
\acro{DASWG}[DASWG]{Data Analysis Software Working Group}
\acro{DBSP}[DBSP]{Double Spectrograph\acroextra{ (instrument on \acs{P200})}}
\acro{DCT}[DCT]{Discovery Channel Telescope}
\acro{DECAM}[DECam]{Dark Energy Camera\acroextra{ (instrument on the Blanco 4\nobreakdashes-m telescope at \acs{CTIO})}}
\acro{DFT}[DFT]{discrete Fourier transform}
\acro{EM}[EM]{electromagnetic}
\acro{FD}[FD]{frequency domain}
\acro{FAR}[FAR]{false alarm rate}
\acro{FFT}[FFT]{fast Fourier transform}
\acro{FIR}[FIR]{finite impulse response}
\acro{FITS}[FITS]{Flexible Image Transport System}
\acro{FLOPS}[FLOPS]{floating point operations per second}
\acro{FOV}[FOV]{field of view}
\acroplural{FOV}[FOV\acrolowercase{s}]{fields of view}
\acro{FTN}[FTN]{Faulkes Telescope North}
\acro{GBM}[GBM]{Gamma-ray Burst Monitor\acroextra{ (instrument on \emph{Fermi})}}
\acro{GCN}[GCN]{Gamma-ray Coordinates Network}
\acro{GMOS}[GMOS]{Gemini Multi-object Spectrograph\acroextra{ (instrument on the Gemini telescopes)}}
\acro{GRB}[GRB]{gamma-ray burst}
\acro{GSL}[GSL]{GNU Scientific Library}
\acro{GTC}[GTC]{Gran Telescopio Canarias}
\acro{GW}[GW]{gravitational wave}
\acro{HAWC}[HAWC]{High\nobreakdashes-Altitude Water \v{C}erenkov Gamma\nobreakdashes-Ray Observatory}
\acro{HCT}[HCT]{Himalayan Chandra Telescope}
\acro{HEALPix}[HEALP\acrolowercase{ix}]{Hierarchical Equal Area isoLatitude Pixelization}
\acro{HEASARC}[HEASARC]{High Energy Astrophysics Science Archive Research Center}
\acro{HETE}[HETE]{High Energy Transient Explorer}
\acro{HFOSC}[HFOSC]{Himalaya Faint Object Spectrograph and Camera\acroextra{ (instrument on \acs{HCT})}}
\acro{HMXB}[HMXB]{high\nobreakdashes-mass X\nobreakdashes-ray binary}
\acroplural{HMXB}[HMXB\acrolowercase{s}]{high\nobreakdashes-mass X\nobreakdashes-ray binaries}
\acro{HSC}[HSC]{Hyper Suprime\nobreakdashes-Cam\acroextra{ (instrument on the 8.2\nobreakdashes-m Subaru telescope)}}
\acro{IACT}[IACT]{imaging atmospheric \v{C}erenkov telescope}
\acro{IIR}[IIR]{infinite impulse response}
\acro{IMACS}[IMACS]{Inamori-Magellan Areal Camera \& Spectrograph\acroextra{ (instrument on the Magellan Baade telescope)}}
\acro{IPAC}[IPAC]{Infrared Processing and Analysis Center}
\acro{IPN}[IPN]{InterPlanetary Network}
\acro{IPTF}[\acrolowercase{i}PTF]{intermediate \acl{PTF}}
\acro{ISM}[ISM]{interstellar medium}
\acro{KAGRA}[KAGRA]{KAmioka GRAvitational\nobreakdashes-wave observatory}
\acro{KDE}[KDE]{kernel density estimator}
\acro{LAT}[LAT]{Large Area Telescope}
\acro{LCOGT}[LCOGT]{Las Cumbres Observatory Global Telescope}
\acro{LHO}[LHO]{\ac{LIGO} Hanford Observatory}
\acro{LIGO}[LIGO]{Laser Interferometer \acs{GW} Observatory}
\acro{llGRB}[\acrolowercase{ll}GRB]{low\nobreakdashes-luminosity \ac{GRB}}
\acro{LLOID}[LLOID]{Low Latency Online Inspiral Detection}
\acro{LLO}[LLO]{\ac{LIGO} Livingston Observatory}
\acro{LMI}[LMI]{Large Monolithic Imager\acroextra{ (instrument on \ac{DCT})}}
\acro{LOFAR}[LOFAR]{Low Frequency Array}
\acro{LSB}[LSB]{long, soft burst}
\acro{LSC}[LSC]{\acs{LIGO} Scientific Collaboration}
\acro{LSO}[LSO]{last stable orbit}
\acro{LSST}[LSST]{Large Synoptic Survey Telescope}
\acro{LTI}[LTI]{linear time invariant}
\acro{MAP}[MAP]{maximum a posteriori}
\acro{MBTA}[MBTA]{Multi-Band Template Analysis}
\acro{MCMC}[MCMC]{Markov chain Monte Carlo}
\acro{MLE}[MLE]{\ac{ML} estimator}
\acro{ML}[ML]{maximum likelihood}
\acro{NED}[NED]{NASA/IPAC Extragalactic Database}
\acro{NSBH}[NSBH]{neutron star\nobreakdashes--black hole}
\acro{NSBH}[NSBH]{\acl{NS}\nobreakdashes--\acl{BH}}
\acro{NSF}[NSF]{National Science Foundation}
\acro{NSNS}[NSNS]{\acl{NS}\nobreakdashes--\acl{NS}}
\acro{NS}[NS]{neutron star}
\acro{OT}[OT]{optical transient}
\acro{P48}[P48]{Palomar 48~inch Oschin telescope}
\acro{P60}[P60]{robotic Palomar 60~inch telescope}
\acro{P200}[P200]{Palomar 200~inch Hale telescope}
\acro{PC}[PC]{photon counting}
\acro{PSD}[PSD]{power spectral density}
\acro{PSF}[PSF]{point-spread function}
\acro{PTF}[PTF]{Palomar Transient Factory}
\acro{QUEST}[QUEST]{Quasar Equatorial Survey Team}
\acro{RAPTOR}[RAPTOR]{Rapid Telescopes for Optical Response}
\acro{REU}[REU]{Research Experiences for Undergraduates}
\acro{RMS}[RMS]{root mean square}
\acro{ROTSE}[ROTSE]{Robotic Optical Transient Search}
\acro{S5}[S5]{\ac{LIGO}'s fifth science run}
\acro{S6}[S6]{\ac{LIGO}'s sixth science run}
\acro{SAA}[SAA]{South Atlantic Anomaly}
\acro{SHB}[SHB]{short, hard burst}
\acro{SHGRB}[SHGRB]{short, hard \acl{GRB}}
\acro{SMT}[SMT]{Slewing Mirror Telescope\acroextra{ (instrument on \acs{UFFO} Pathfinder)}}
\acro{SNR}[S/N]{signal\nobreakdashes-to\nobreakdashes-noise ratio}
\acro{SSC}[SSC]{synchrotron self\nobreakdashes-Compton}
\acro{SDSS}[SDSS]{Sloan Digital Sky Survey}
\acro{SED}[SED]{spectral energy distribution}
\acro{SN}[SN]{supernova}
\acroplural{SN}[SN\acrolowercase{e}]{supernova}
\acro{SNIcBL}[\acs{SN}\,I\acrolowercase{c}\nobreakdashes-BL]{broad\nobreakdashes-line Type~Ic \ac{SN}}
\acroplural{SNIcBL}[\acsp{SN}\,I\acrolowercase{c}\nobreakdashes-BL]{broad\nobreakdashes-line Type~Ic \acp{SN}}
\acro{SVD}[SVD]{singular value decomposition}
\acro{TAROT}[TAROT]{T\'{e}lescopes \`{a} Action Rapide pour les Objets Transitoires}
\acro{TDOA}[TDOA]{time delay on arrival}
\acroplural{TDOA}[TDOA\acrolowercase{s}]{time delays on arrival}
\acro{TD}[TD]{time domain}
\acro{TOA}[TOA]{time of arrival}
\acroplural{TOA}[TOA\acrolowercase{s}]{times of arrival}
\acro{TOO}[TOO]{target\nobreakdashes-of\nobreakdashes-opportunity}
\acroplural{TOO}[TOO\acrolowercase{s}]{targets of opportunity}
\acro{UFFO}[UFFO]{Ultra Fast Flash Observatory}
\acro{UHE}[UHE]{ultra high energy}
\acro{UVOT}[UVOT]{UV/Optical Telescope\acroextra{ (instrument on \emph{Swift})}}
\acro{VHE}[VHE]{very high energy}
\acro{VLA}[VLA]{Karl G. Jansky Very Large Array}
\acro{VLT}[VLT]{Very Large Telescope}
\acro{WAM}[WAM]{Wide\nobreakdashes-band All\nobreakdashes-sky Monitor\acroextra{ (instrument on \emph{Suzaku})}}
\acro{WCS}[WCS]{World Coordinate System}
\acro{WSS}[w.s.s.]{wide\nobreakdashes-sense stationary}
\acro{XRF}[XRF]{X\nobreakdashes-ray flash}
\acroplural{XRF}[XRF\acrolowercase{s}]{X\nobreakdashes-ray flashes}
\acro{XRT}[XRT]{X\nobreakdashes-ray Telescope\acroextra{ (instrument on \emph{Swift})}}
\acro{ZTF}[ZTF]{Zwicky Transient Facility}
\end{acronym}

\begin{figure}
    \centering
    \includegraphics[width=\columnwidth]{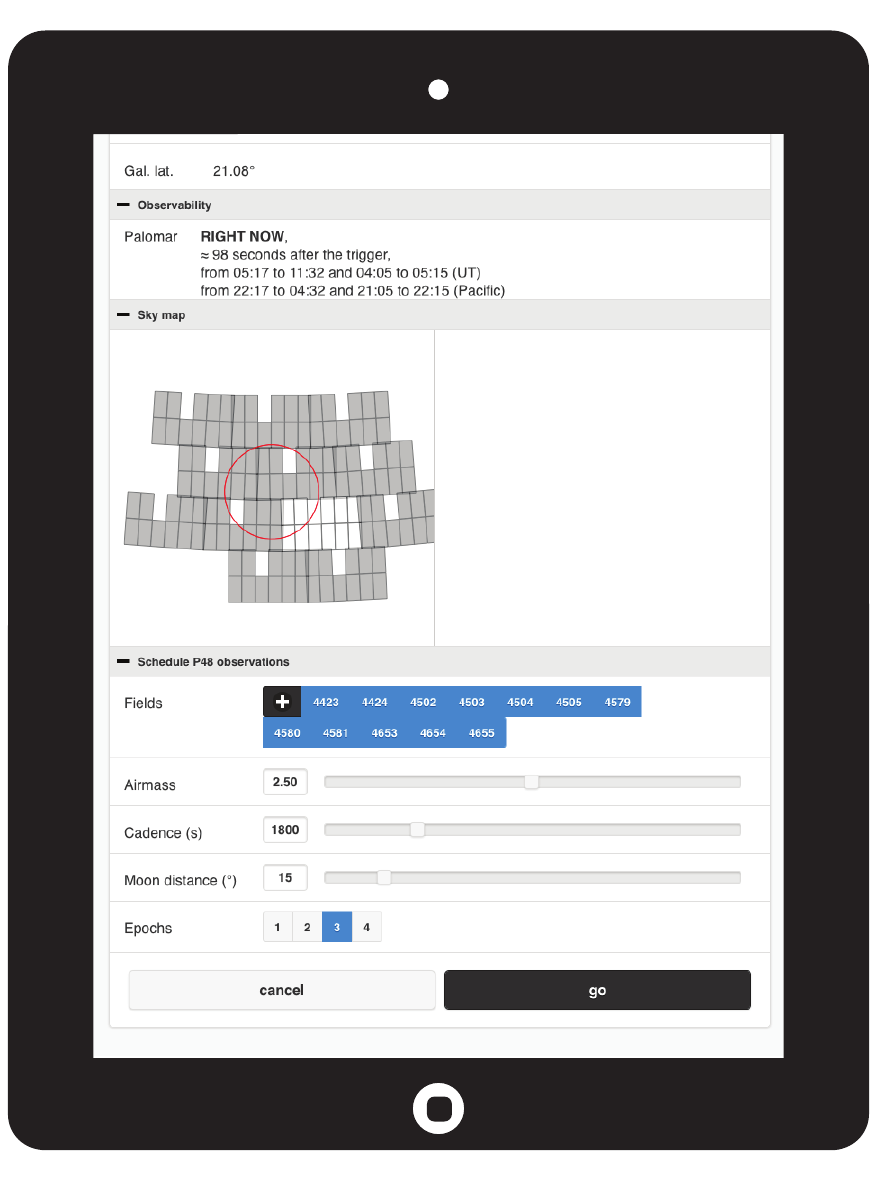}
    \caption[Screenshot of \acs{IPTF} \acs{TOO} Marshal]{\label{fig:screenshot}Screenshot of the \ac{IPTF} \ac{TOO} Marshal shortly after a \emph{Fermi} \ac{GBM} detection. At this stage, the application presents the recommended \ac{P48} fields, the time window of observability, and the history of \ac{GCN} notices and circulars related to the trigger. It gives the human participants the option to customize the \ac{P48} sequence by adding or removing \ac{P48} fields and tuning the airmass limit, cadence, or number of images.
}
\end{figure}

\begin{figure}
    \centering
    \includegraphics[width=\columnwidth]{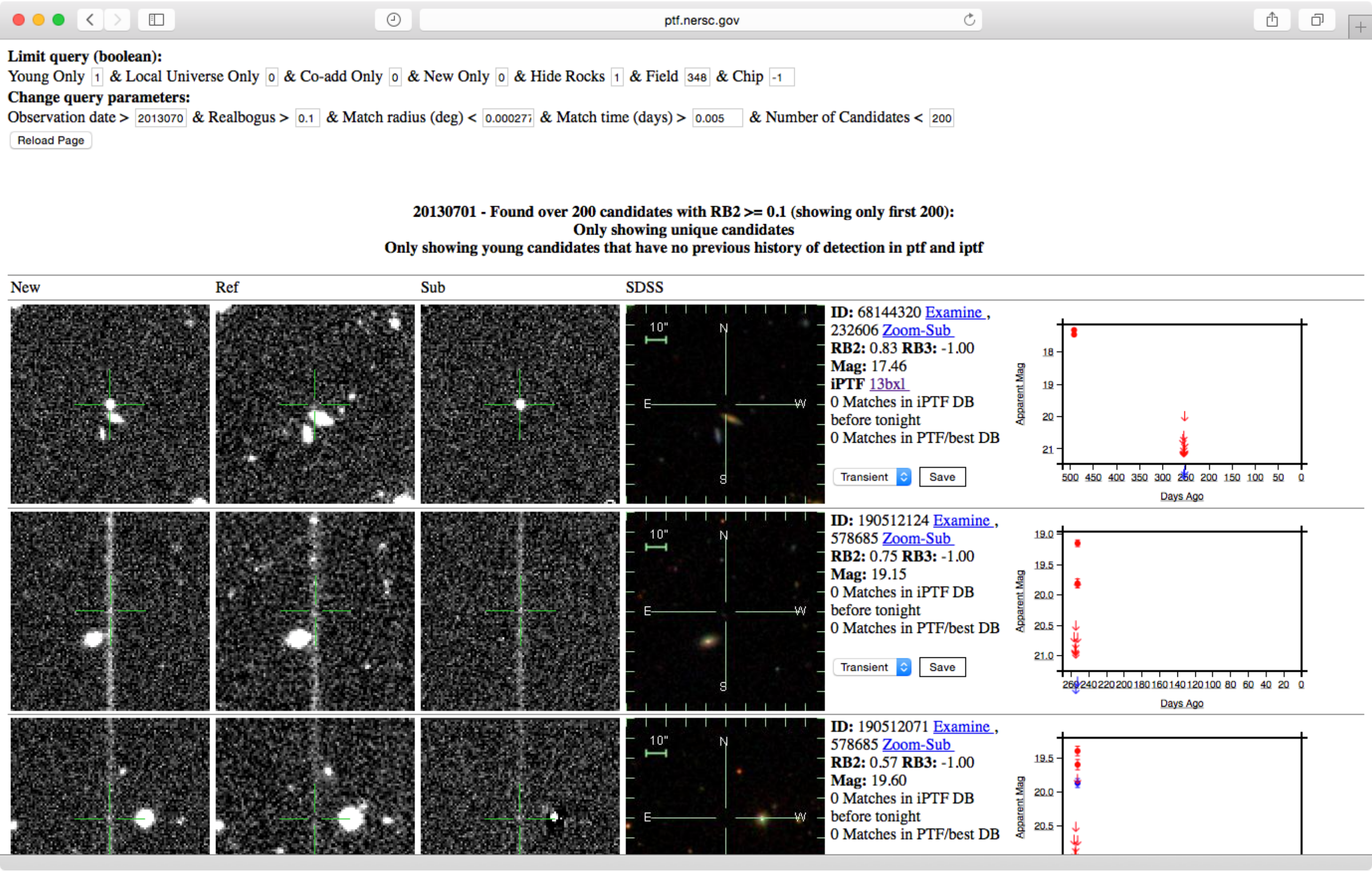}
    \caption[Screenshot of \acs{IPTF} Treasures portal]{\label{fig:treasures}Screenshot of the Treasures portal, showing new, reference, subtraction, and archival \ac{SDSS} images as well as \ac{P48} light curves. This page is for the date and field containing \ac{GRB}~130702A~/~iPTF13bxl.}
\end{figure}

\begin{figure}
    \centering
    \includegraphics[width=\columnwidth]{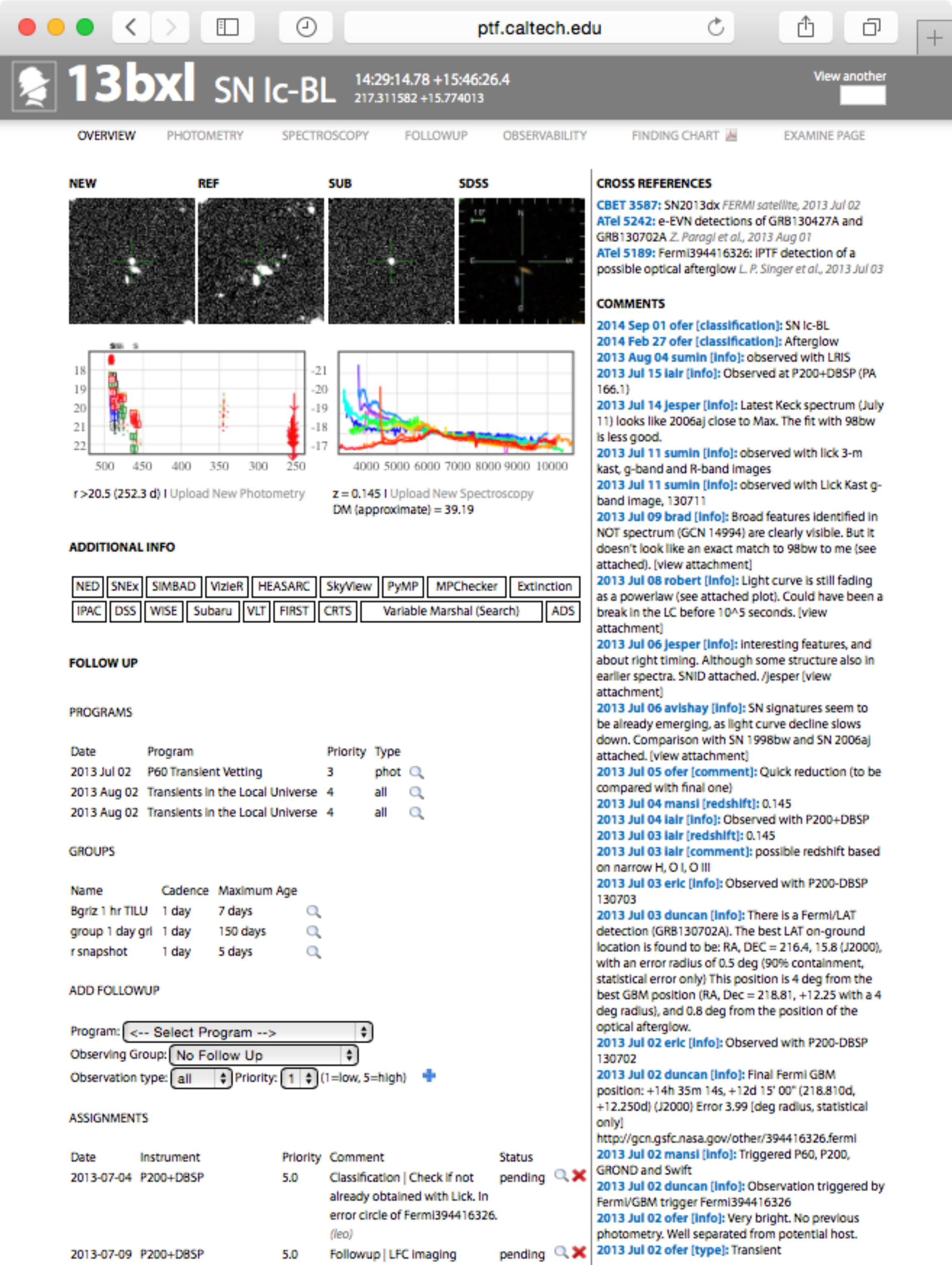}
    \caption[Screenshot of \acs{IPTF} Transient Marshal]{\label{fig:transient-marshal}Screenshot of the \ac{IPTF} Transient Marshal, showing \ac{GRB}~130702A~/~iPTF13bxl.}
\end{figure}

\bibliographystyle{apj}
\bibliography{ms,gcn}

\end{document}